\documentstyle[prb,aps]{revtex}
\begin{document}
\advance\textheight by 0.5in
\advance\topmargin by -0.25in
\draft

\twocolumn[\hsize\textwidth\columnwidth\hsize\csname@twocolumnfalse%
\endcsname

\preprint{NSF--ITP--95--25, cond-mat/9503084}

\title{ {\hfill\normalsize NSF--ITP--95--25, cond-mat/9503084\medskip\\}
Quantum Smectic and Supersolid Order in Helium Films and Vortex Arrays}
\author{Leon Balents}
\address{Institute for Theoretical Physics, University of California,
Santa Barbara, CA 93106--4030}
\author{David R. Nelson}
\address{Lyman Laboratory of Physics, Harvard University, Cambridge, MA 02138}

\date{\today}
\maketitle

\begin{abstract}
A flux liquid can condense into a smectic crystal in a pure layered
superconductor with the magnetic field oriented nearly parallel to the
layers.  Similar order can arise in low temperature $^4$He films with
a highly anisotropic periodic substrate potential.  If the smectic
order is commensurate with the layering, this periodic array is {\sl
stable} to quenched random point--like disorder.  By tilting and
adjusting the magnitude of the applied field, both incommensurate and
tilted smectic and crystalline phases are found for vortex arrays.
Related variations are possible by changing the chemical potential in
the helium system.  We discuss transport near the second order smectic
freezing transition, and show that permeation modes in superconductors
lead to a small non--zero resistivity and a large but finite tilt
modulus in the smectic crystal.  In helium films, the theory predicts
a nonzero superfluid density and propagating third sound modes,
showing that the quantum smectic is always simultaneously crystalline
and superfluid.
\end{abstract}
\pacs{PACS: 74.60.Ge,74.40.+k}
]

\section{Introduction}
\label{Introduction}

The discovery of high temperature superconductors, with their broad
fluctuation regime, has emphasized the inadequacy of the conventional
mean--field description of critical behavior.\cite{FluctExpts}\ Early
attempts to apply field theoretic methods to the Ginzburg--Landau (GL)
free energy, however, are of limited applicability in low
dimensions.\cite{HLM,NBT}\

Instead, superconducting fluctuations in low dimensions are now
understood in terms of vortices, which emerge as the low energy
degrees of freedom of the Ginzburg--Landau theory.  The phases of the
superconductor within this picture are analogous to states of
conventional matter, except that they are composed of flux lines
instead of molecules.  In fact, because the vortices are extended
objects, the system most closely resembles a collection of {\sl
quantum} bosons.\cite{NelsonSeung}\ As the thickness of the
superconductor approaches infinity, the effective ``temperature'' of
this bosonic system goes to zero, and interesting strongly--correlated
phases can emerge.

One difference between the flux line array and true assemblies of bosons is
that in the former, the effects of the embedding medium are more often
dramatic.  Indeed, without careful preparation, most high
temperature superconducting samples are dominated by internal defects
which tend to disorder the vortex array.  Only at reasonably high
temperatures, when the fluxons are best described as a liquid, can the
effects of these random pinning centers be neglected.\cite{NlD}\  At
lower temperatures, disorder may induce subtle types of glassy
order,\cite{VGtheory,BGtheory,VGexperiments,BGexperiments}\ or simply
force the system to remain a liquid with very sluggish dynamics.\cite{Young}\

The layered structure of the copper--oxide materials itself provides a
non--random source of pinning.\cite{intrinsicrefs}\  At low temperatures,
the c-axis coherence length $\xi_{c0} \approx 4 \AA \lesssim s
\approx 12 \AA$, where $s$ is the lattice constant in this direction.
Vortex lines oriented in the ab plane are attracted to the regions of
low condensate electron density between the CuO$_2$ layers.  Such a
periodic potential for true two--dimensional bosons could be induced
by an anisotropically corrugated substrate, possibly leading to the
observation of the effects described here in He$^4$ films.

Previous work on intrinsically pinned vortices has focused on the low
temperature fluctuationless regime, in which the vortices form a
pinned elastic solid.  Near $T_c$, however, when thermal fluctuations
are important, entirely different phases can exist.  These thermally
fluctuating states are particularly interesting experimentally because
hysteretic effects are weak and equilibrium transport measurements are
more easily performed than at low temperatures.  Our research is
motivated by the recent experimental work of Kwok et.
al.\cite{Kwok}\, who observed a continuous resistive transition in
${\rm YBa_2Cu_3O_7}$ for fields very closely aligned ($\theta <
1^\circ$) to the ab plane.  A preliminary version of our results
appeared in Ref.\onlinecite{BNprl}.

To explain the experiments, the interplay between inter--vortex
interactions and thermal fluctuations must be taken into account in an
essential way.  The experiments of Ref.\onlinecite{Kwok}\ seem to rule
out conventional freezing, which is first order in all known
three--dimensional cases.  The additional observation of a strong
first order freezing transition for $\theta \gtrsim 1^\circ$ suggests
that point disorder is relatively unimportant at these elevated
temperatures (strong point disorder would destroy a first order
freezing transition).  In addition, an attempted fit of the data to a
dynamical scaling form yielded exponents inconsistent with vortex or
Bose glass values.\cite{Kwok}\  Instead, we postulate freezing into an
intermediate ``smectic'' phase between the high temperature flux
liquid and a low temperature crystal/glass.  Such smectic freezing, as
discussed by de Gennes for the nematic--smectic A
transition,\cite{DeGennes}\ can occur via a continuous transition in
three dimensions.  The vortex smectic state is richer than its liquid
crystal counterpart, however, for two reasons.  First, the existence
of a periodic embedding medium (i.e.  the crystal lattice) in the
former leads to commensurability effects not present in the liquid
crystal.\cite{Villainreview}\ In addition, the connectedness of flux
lines leads to constraints with no analog for pointlike molecules.  As
we show below, the onset of smectic order should be accompanied by a
steep drop in the resistivity and a rapid increase in the tilt modulus
for fields which attemp to tip vortices out of the CuO$_2$ planes --
see Fig.\ref{qualfig}.

\begin{figure}
\caption{Schematic structure functions in the (a) liquid, (b) smectic,
and (c) solid phases.  For simplicity, we have illustrated the case
for a square lattice.}
\label{sffig}\end{figure}

The smectic phase may also be distinguished experimentally using neutron
scattering, which measures the Fourier transform of the magnetic field
two--point correlation function (see Fig.\ref{sffig}).  We assume a
magnetic field along the $y$--axis and CuO$_2$ layers perpendicular to
${\bf \hat{z}}$.  The vortex liquid structure function shows the usual
diffuse liquid rings, as well as delta--function Bragg peaks at $q_z =
2\pi n/s$ (for integral $n$), representing the ``imposed'' vortex
density oscillations from the CuO$_2$ layers.  On passing to the
smectic state, additional peaks develop at wavevectors $q_z = 2\pi
n/a$, interlacing between those already present in the liquid.  The
new peaks represent the broken symmetry associated with preferential
occupation of a periodic {\sl subset} of the layers occupied by the
vortices in the liquid.  At lower temperatures in the vortex solid,
further peaks form for $q_x \neq 0$, producing the full reciprocal
lattice of a two--dimensional crystal.

Our analysis leads to the phase diagrams shown in
Fig.\ref{phasediagramfig}.  Upon lowering the temperature for $H_c =
0$ and a commensurate value of $H_b$, the vortex liquid (L) freezes
first at $T_s$ into the pinned smectic (S) state, followed by a second
freezing transition at lower temperatures into the true vortex crystal
(X).  When $H_c \neq 0$, tilted smectic (TS) and crystal (TX) phases
appear.  The TS--L and TX--TL transitions are XY--like, while the
TS--S and TX--X phase boundaries are commensurate--incommensurate
transitions (CITs).\cite{Villainreview}\ At larger tilts, the TX--TS
and TS--L phase boundaries merge into a single first order melting line.  As
$H_b$ is changed, incommensurate smectic (IS) and crystal (IX) phases
appear, again separated by CITs from the pinned phases, and an XY
transition between the IS and L states.  If the low--$H_b$ commensurate
smectic (S) phase corresponds to, say, 5 CuO$_2$ plane periodicities
per vortex layer, the high--$H_b$ S phase will represent a state with
4 CuO$_2$ plane periodicities per sheet.  We show that the commensurate
smectic order along the $c$ axis is {\sl stable} to weak point
disorder, in striking contrast to the triangular flux lattice which
appears for fields aligned with the c axis.\cite{LO} .  This stability
should {\sl increase} the range of smectic behavior relative to the
(unstable) crystalline phases when strong point disorder is present.

\begin{figure}
\caption{Phase diagrams in the (a) $H_c$--$T$, and (b) $H_b$--$T$
planes for the disorder--free vortex system.  The fuzzy lines indicate
first order transitions.  In (b) two possible topologies are shown
connecting different commensurate states.}
\label{phasediagramfig}\end{figure}

The high temperature flux liquid (for fields in the ab plane) is of
some interest in its own right.  Consider first one vortex line
wandering along the $y$--axis, as shown schematically in
Fig.\ref{wanderfig}.  This line is subject only to thermal
fluctuations and a periodic pinning potential along the $z$--axis,
provided by the CuO$_2$ planes.\cite{cuprateref}\  If thermal
fluctuations are ignored, the vortex acts like a rigid rod, and will
be localized in one of the potential minima.\cite{xdelocref}\  This
localization assumption, however, is {\sl always} incorrect in the
presence of thermal fluctuations, provided the sample is sufficiently
large in the $y$--direction.  As $L_y \rightarrow \infty$, the
statistical mechanics of this single wandering line random walking in
directions perpendicular to $y$ leads inevitably to equal
probabilities that the vortex is in {\sl any} of the many possible
minima along ${\bf \hat{z}}$.

On a more formal level, this probability distribution $P(z)$ is given
by the square of the ground state wave function of the Schr\"odinger
equation in a periodic potential -- see section \ref{liquidsec}\
below.  The jumps shown in Fig.\ref{wanderfig}\ across CuO$_2$ planes
are represented by quantum mechanical tunneling in imaginary time.
According to Bloch's theorem, this tunneling leads to $P(z) =
|\psi_{k=0}(z)|^2$, where the Bloch states in general have the form
$\psi_k(z) = \exp(ikz)u(z)$, with $u(z)$ a function with the
periodicity of the pinning potential.  The resulting probability
distribution is shown schematically on the right side of
Fig.\ref{wanderfig}.

\begin{figure}
\caption{Wandering of a single flux line (solid curve), leading to an
extended probability distribution $P(z)$ given by a $k=0$ Bloch
wavefunction.  Other vortex trajectories (represented by the dashed
curve) will generate similar probability distributions, unless
interactions lead to crystalline or smectic order.}
\label{wanderfig}
\end{figure}

Now suppose an additional line is added to the system.  As suggested
by the trajectory of the dashed curve in Fig.\ref{wanderfig}, it too
will wander from plane to plane.  Although the two flux lines interact
repulsively, they can wander and still avoid each other by using the
$x$--coordinate or by never occupying the same minimum at the same
value of the ``imaginary time'' coordinate $y$.  Thus {\sl both} flux
lines generate a delocalized probability distribution and occupy the
same $k=0$ Bloch state.  At high temperatures or when the lines are
dilute, we expect for similar reasons {\sl macroscopic occupation} of
the $k=0$ Bloch state in the equivalent boson many body quantum
mechanics problem, similar to Bose--Einstein condensation.  In this
sense, the flux liquid is indeed a ``superfluid.''  The presence of
numerous ``kinks'' in the vortex trajectories insures a large tilt
response for fields along ${\bf \hat{z}}$ and a large resistivity for
currents along ${\bf \hat{x}}$.  The various
symmetry--breaking crystalline or smectic states which appear at low
temperatures or higher densities arise because of the localizing
tendency of the interactions.  The density of kinks is greatly reduced
in these phases.

The remainder of the paper is organized as follows.  In section
\ref{modelsec}, several models are introduced which will be used to
analyze the layered superconductor.  Sections \ref{liquidsec}\ and
\ref{CrystalPhaseSection}\ discuss the effect of intrinsic pinning on
the liquid and crystal phases, respectively, and show how smectic
ordering is encouraged on approaching the intermediate regime from
these two limits.  A Landau theory for the liquid--smectic transition
is introduced in section \ref{criticalsection}, and the critical
behavior is determined within this model.  The
nature of the commensurate smectic phase itself is explored in section
\ref{smecticsection}, through a computation of the response functions.
In section \ref{supersolidsection}, it is shown to have ``supersolid''
order, similar to the supersolid crystal phase recently proposed at
high magnetic fields along the ${\bf\hat{c}}$ axis.\cite{FNF}\  The
additional phases which arise for large incommensurate fields are
described in section \ref{ICphases}.  Section \ref{disordersection}\
details the modifications of the phase diagram when weak point
disorder is present, and, in particular, demonstrates the stability of
the smectic state.  Concluding remarks and the
implications of these results for helium films on periodically ruled
substrates are presented in section \ref{conclusionsection}.

\section{Models}
\label{modelsec}

At a fundamental level (within condensed matter physics), layered
superconductors may be modeled as a positively charged ionic
background and a collection of conduction electrons, which can pair
via the exchange of phonons, excitons, magnons, etc.  Since such a
microscopic theory of high temperature superconductors is lacking, we
must resort to more phenomenological methods.  There are,
nevertheless, a variety of differing levels of description available,
several of which will be used in the remainder of this paper.

\subsection{Static Models}

The most basic of our models is the familiar Ginzburg--Landau theory, which is
an expansion of the free energy of the
superconductor in powers of the order parameter field $\Psi_{\rm GL}$,
\begin{eqnarray}
F_{\rm GL} & = & F_{\rm SC} + F_{\rm EM}, \nonumber \\
F_{\rm SC} & = & \int d^3{\bf r} \bigg\{ \sum_\mu{\hbar^2 \over
{2m_\mu}}\left|D_\mu\Psi_{\rm GL}\right|^2 + \alpha|\Psi_{\rm GL}|^2
\nonumber\\
& & +
{\beta \over 2}|\Psi_{\rm GL}|^4\bigg\}, \nonumber\\
F_{\rm EM} & = &\int d^3{\bf r} {1 \over {8\pi}}|\bbox{\nabla} \times {\bf
A} - {\bf H}|^2,
\label{GLfreeenergy}
\end{eqnarray}
where $D_\mu \equiv \partial_\mu - 2\pi A_\mu/\phi_0$, $\phi_0 =
hc/2e$ is the flux quantum.  Here ${\bf H}$ is the applied external
magnetic field.  To establish notation, we choose $x$, $y$, and $z$
respectively along the $a$, $b$, and $c$ axes of the underlying
cuprate crystal.  The diagonal components of the effective mass tensor
are then $m_x = m_y \equiv m$ and $m_z = M$.  For convenience in what
follows, we also define the anisotropy ratio $\gamma \equiv \sqrt{m/M}
= \lambda_{ab}/\lambda_c = \xi_c/\xi_{ab} \ll 1$.

Eq.\ref{GLfreeenergy}\ provides a powerful means of understanding
superconducting behavior, including the effects of anisotropy.  In layered
materials, however, the theory must be modified to allow for coupling of
the superconducting order to the crystalline lattice.  One such
description, in which the superconductor is regarded as a stack of
Josephson--coupled layers, is the Lawrence--Doniach
model.  For our purposes, however, it is sufficient to consider a
``soft'' model for the lattice effects, in which the coupling $\alpha$ is
allowed to be a periodic function of $z$ with period $s$ equal to the
periodicity of the copper--oxide planes.

Because the high temperature superconductors are strongly type II, it is
appropriate to use London theory over a large range of the phase
diagram.  In this limit, variations of the magnitude
of the order parameter are confined to a narrow region within
the core of each vortex.  Because the resulting London equations
are linear, a complete solution can be obtained for the free energy of an
arbitrary vortex configuration.\cite{Brandt}\ For our purposes, it is
sufficient to consider an approximate form in which the tilt moduli
are local and the interactions between vortices occur at equal
$y$,\cite{localnote}\
\begin{eqnarray}
F_{\rm London} & & = \sum_i \int \!\! dy \, {\tilde{\epsilon}_\parallel \over
2} \left| {{dx_i(y)} \over {dy}}\right|^2 +
{\tilde{\epsilon}_\perp \over 2} \left| {{dz_i(y)} \over
{dy}}\right|^2 \nonumber \\ & & \!\!\! - V_{\rm P}[z_i(y)] + \sum_{i,j} {1
\over 2} \int dy V[{\bf r}_{\perp i}(y) - {\bf r}_{\perp j}(y)],
\label{londonvortexfreeenergy}
\end{eqnarray}
where $V[{\bf r}_\perp] \approx
2\gamma\epsilon_0K_0(\sqrt{x^2+z^2/\gamma^2}/\lambda_{ab})$
($K_0(r)$ is a modified Bessel function), $\epsilon_0 =
(\phi_0/4\pi\lambda_{ab})^2$ ), and the stiffness constants obtained
from anisotropic Ginzburg--Landau (GL) theory are
$\tilde{\epsilon}_\parallel =
\epsilon_0\gamma$ and $\tilde{\epsilon}_\perp = \epsilon_0/\gamma$
 (see, e.g. Ref.\onlinecite{BGL}).  $V_{\rm P}[z]$ is a periodic
potential ($V_{\rm P}[z+s] = V_{\rm P}[z]$) taking into account the
effects of the layering.

Either Eq.\ref{GLfreeenergy}\ or Eq.\ref{londonvortexfreeenergy}\ may be
used at finite temperature, by calculating the partition function
\begin{equation}
Z = {\rm Tr} e^{-F/k_{\rm B}T},
\label{partitionfunction}
\end{equation}
where the trace is a functional integral over $\Psi_{\rm GL}$ or the set
of vortex trajectories $\{ {\bf r}_i(z) \}$, for the Ginzburg--Landau and
London limits,
respectively.

In the London case, this trace may be formally performed by
recognizing Eq.\ref{partitionfunction}\ as mathematically identical to
the Feynmann path integral for the first quantized imaginary time
Green's function of interacting bosons.  Within this boson
analogy,\cite{NelsonSeung}\ the Green's function may also be
calculated using a coherent state path integral representation.  The
``action'' for these bosons is
\begin{eqnarray}
\tilde{S}_{\rm boson} & & =  \int d^3{\bf r} \psi^\dagger
 (T\partial_y - {{T^2} \over {2\tilde{\epsilon}_\perp}}\partial_z^2 -
{{T^2} \over {2\tilde{\epsilon}_\parallel}}\partial_x^2 - \mu)\psi
\nonumber \\
& & - \int d^3{\bf r} V_{\rm P}(z) n({\bf r}) \nonumber \\
& & \!\! + \int \! d^2{\bf r_\perp} d^2{\bf r'_\perp} dy {1 \over
2}
V({\bf r_\perp\! -\! r'_\perp})n({\bf r}_\perp,y)n({\bf
r'_\perp},y).
\label{anisotropicbosonaction}
\end{eqnarray}
where $\psi$ is the complex coherent state boson field, and $n({\bf r}) =
\psi^\dagger({\bf r})\psi({\bf r})$.  It is convenient to
rescale $x \rightarrow
(\tilde{\epsilon}_\perp/\tilde{\epsilon}_\parallel)^{1/2}x$ and $z
\rightarrow
(\tilde{\epsilon}_\parallel/\tilde{\epsilon}_\perp)^{1/2}z$ to obtain
the isotropic Laplacian $\nabla_\perp^2 \equiv \partial_x^2 +
\partial_z^2$.  Eq.\ref{anisotropicbosonaction}\ becomes
\begin{eqnarray}
S_{\rm boson} & & = \int d^3{\bf r} \left[\psi^\dagger (T\partial_y - {{T^2}
\over {2\tilde{\epsilon}}}\nabla_\perp^2 - \mu)\psi -\tilde{V}_{\rm
P}(z) n({\bf r})\right] \nonumber \\ & &\!\! + {1 \over 2}\int\!
d^2{\bf r_\perp} d^2{\bf r'_\perp} dy \tilde{V}({\bf r_\perp\! -\!
r'_\perp})n({\bf r}_\perp,y)n({\bf r'_\perp},y),
\label{bosonaction}
\end{eqnarray}
with $\tilde{\epsilon} \equiv
\sqrt{\tilde{\epsilon}_\parallel\tilde{\epsilon}_\perp}$ and the
rescaled potentials $\tilde{V}_{\rm P}(z) = V_{\rm
P}(z\gamma^2)$ and $\tilde{V}({\bf r}_\perp) = V(x/\gamma^2
,z\gamma^2)$.  The action is used via
\begin{equation}
{\cal Z} = \int [d\psi^\dagger][d\psi] e^{-S/k_{\rm B}T}
\label{grandcanonical}
\end{equation}
to calculate the grand canonical partition function ${\cal Z}$ at
chemical potential $\mu$ per unit length of vortex line.
Eqs.\ref{grandcanonical}\ and \ref{bosonaction}\ may also be obtained
directly froma limiting case of Eq.\ref{GLfreeenergy}\ via a duality
mapping.\cite{duality}\

\subsection{Dynamical Models}
\label{hydromodel}

Calculation of dynamical response functions such as the resistivity
requires a model for the time dependence of the superconductor.  We
will do this within the London framework, treating the vortex lines as
the dynamical degrees of freedom.  In the overdamped limit, the
appropriate equation of motion is then
\begin{equation}
\Gamma \dot{{\bf r}}_{\perp i}(y) = - {{\partial F_{\rm London}} \over
{\partial{\bf r}_{\perp i}(y)}} + {\bf f}_i,
\label{FLeom}
\end{equation}
where $\Gamma$ is a damping constant, and ${\bf f}_i$ is the force on
the $i^{\rm th}$ flux line, including both external forces and random thermal
noise.

Eq.\ref{FLeom}\ is useful in describing the properties of small
numbers of vortices.  To understand the bulk behavior of dense phases,
however, we need an extensive theory.  Such a model for the liquid
phase was constructed on physical and symmetry grounds in
Ref.\onlinecite{MNhydro}\ in the hydrodynamic limit.  Because we
intend to go beyond simple linearized hydrodynamics,
we require some knowledge of the non--linear form of the bulk equations
of motion.

We proceed by first defining the hydrodynamic fields
\begin{eqnarray}
n({\bf r}) & = & \sum_i \delta[{\bf r}_\perp - {\bf r}_{\perp i}(y)],
\label{ndef} \\
\bbox{\tau}({\bf r}) & = & \sum_i \delta[{\bf r}_\perp - {\bf r}_{\perp
i}(y)] {{d{\bf r}_{\perp i}(y)} \over {dy}}, \label{taudef} \\
{\bf f}({\bf r}) & = & \sum_i \delta[{\bf r}_\perp - {\bf r}_{\perp
i}(y)] {\bf f}_i. \label{fdef}
\end{eqnarray}
Conservation of magnetic flux is embodied in the constraint
\begin{equation}
\partial_y n + \bbox{\nabla}_\perp \cdot \bbox{\tau} = 0.
\label{constraint}
\end{equation}
Eq.\ref{londonvortexfreeenergy}\ may be rewritten as a function of ${\bf
n}$ and $\bbox{\tau}$, formally
\begin{equation}
F_{\rm London}[\{ {\bf r}_{\perp i}(y) \}] = F[n({\bf
r}),\bbox{\tau}({\bf r})].
\label{Fdef}
\end{equation}
Specific forms for $F[n,\bbox{\tau}]$ will be used as needed.

Eq.\ref{FLeom}\ completely specifies the dynamics of $n$ and
$\bbox{\tau}$.  Differentiating Eqs.\ref{ndef}\ and \ref{taudef}\ then leads
immediately to hydrodynamic equations of motion.
Details are given in appendix \ref{dynamicsappendix}.  One finds
\begin{eqnarray}
\partial_t n + \bbox{\nabla}_\perp\cdot{\bf j}_v & = & 0,
\label{continuity} \\
\partial_t\tau_\alpha + \partial_\beta j_{\beta\alpha} & = & \partial_y
j_{v,\alpha}, \label{taucontinuity}
\end{eqnarray}
where the density and tangent currents are
\begin{eqnarray}
\Gamma {\bf j}_v & = & -n\bbox{\nabla}_\perp{{\delta F} \over
{\delta n}} + n\partial_y{{\delta F} \over {\delta\bbox{\tau}}}
-\tau_\alpha\bbox{\nabla}_\perp{{\delta F} \over {\delta\tau_\alpha}}
+ n{\bf f},
\label{constitutive} \\
\Gamma j_{\beta\alpha} & = & -\tau^\beta\partial_\alpha{{\delta F}
\over {\delta n}} + \tau^\beta\partial_y {{\delta F} \over {\delta
\tau_\alpha}} - \tau^{\beta\gamma}\partial_\alpha {{\delta F} \over
{\delta\tau_\gamma}} + \tau^\beta f^\alpha \nonumber \\ & & - (\beta
\leftrightarrow \alpha),
\label{tauconstitutive}
\end{eqnarray}
where
\begin{equation}
\tau^{\beta\gamma}({\bf r}) \equiv \sum_i \delta[{\bf r}_\perp - {\bf
r}_{\perp i}(y)]{{dx^\beta_i} \over {dy}}{{dx^\gamma_i} \over {dy}}.
\label{tau2def}
\end{equation}

Because the constitutive relation for the tangent current includes
$\tau^{\beta\gamma}$, Eqs.\ref{continuity}--\ref{tau2def}\ do not form
a closed set.  Vortex hydrodynamics, however, leads us to expect that
$n$ and $\bbox{\tau}$ provide a complete long--wavelength description
of the system.  We therefore adopt the truncation scheme
$\tau^{\beta\gamma} \rightarrow \langle \tau^{\beta\gamma} \rangle$
(i.e. $\tau^{\beta\gamma}$ is preaveraged at equilibrium).  Averaging
via Eq.\ref{londonvortexfreeenergy} gives
\begin{equation}
\tau^{\beta\gamma} \rightarrow k_{\rm B}T n \left( {{\delta^{\beta
x}\delta^{\gamma x}} \over {\tilde{\epsilon}_\parallel}} + {{\delta^{\beta
z}\delta^{\gamma z}} \over {\tilde{\epsilon}_\perp}}  \right).
\label{preaveraging}
\end{equation}
To complete the dynamical description, we must specify $\Gamma$ and
${\bf f}$.  Matching to the liquid hydrodynamics of
Ref.\onlinecite{MNhydro}\ relates $\Gamma$
to the Bardeen--Stephen friction coefficient $\gamma_{\rm
BS} = n_0\Gamma$, and gives the driving force
\begin{equation}
{\bf f} = {{\phi_0} \over c}{\bf J \wedge \hat{y}} +
\bbox{\eta},
\label{forces}
\end{equation}
Here ${\bf J}$ is the applied transport current density and
$\bbox{\eta}({\bf r})$ is a random thermal noise.

\section{Intrinsic Pinning in the Vortex Liquid}
\label{liquidsec}

To better understand the interplay of thermal fluctuations and
layering of the superconductor, it is useful to first consider the
behavior at high temperatures in the liquid state.  As discussed by
Marchetti,\cite{MCMhydro}\ the physics of a single vortex line in the
liquid is well described by a hydrodynamic coupling to the motion of
other vortices.  In the dense limit, this medium is approximately
uniform, and does not significantly effect the wandering of an
isolated vortex.  To estimate the effects of intrinsic pinning, it is
thus appropriate to consider a single vortex line, oriented along the
$a-b$ plane.

For $T \gtrsim
80K$, $\xi_c \approx
\xi_{c0}(1-T/T_c)^{-1/2} \gtrsim s$.  In this limit, the copper--oxide
planes act as a smooth periodic potential on the vortex.  The magnitude
of this potential per unit length is\cite{BLO}
\begin{equation}
U_p \approx 5\times 10^2 \epsilon_0\gamma\left({{\xi_c}
\over s}\right)^{5/2} e^{-15.8 \xi_c/s}.
\label{upinning}
\end{equation}

For a single vortex, Eq.\ref{londonvortexfreeenergy}\ reduces to
\begin{equation}
F_{\rm v} = \int_0^L \!\!\! dy \!\!\left\{ {{\tilde{\epsilon}_\parallel} \over
2}\left|{{dx} \over {dy}}\right|^2\!\! + {{\tilde{\epsilon}_\perp} \over
2}\left|{{dz} \over {dy}}\right|^2\!\! - V_{\rm P}(z) \right\}.
\label{singlevortexfreeenergy}
\end{equation}
The periodic potential $V_{\rm P}[z] = U_p f_p [z/s]$, where
$f_p(u) = f_p(u + 1)$ is a smooth periodic function with magnitude of
order unity.

The $x$--displacement decouples in Eq.\ref{singlevortexfreeenergy},
and may be integrated out to yield
\begin{equation}
\langle [x(y)-x(0)]^2 \rangle \sim D_x y,
\label{xdiffusion}
\end{equation}
with $D_x = k_{\rm B}T/\tilde{\epsilon}_\parallel$.  The
$z$--dependent part of
Eq.\ref{singlevortexfreeenergy}\ is identical to the Euclidean action
of a quantum particle of mass $\tilde{\epsilon}_\perp$ in a
one--dimensional periodic potential $V_{\rm P}(z)$, with $y$ playing
the role of imaginary time.  The single flux line partition function,
\begin{equation}
{\cal Z}_1 = \int [dz(y)] e^{-F_{\rm v}[z]/k_{\rm B}T},
\label{singleFLpartitionfunction}
\end{equation}
with fixed endpoints, maps to the Euclidean Green's function for the
particle, with $k_{\rm B}T$ replacing $\hbar$.

In the quantum--mechanical analogy, the particle tunnels between
adjacent minima of the pinning potential, leading, as discussed in the
Introduction, completely delocalized Bloch wavefunctions even for {\sl
extremely} strong pinning.  The ``time'' required for this tunneling
maps to the distance, $L_{\rm kink}$, in the $y$--direction between
kinks in which the vortex jumps across one CuO$_2$ layer.  The WKB
approximation gives
\begin{equation}
L_{\rm kink} \sim s\sqrt{{{\tilde{\epsilon}_\perp} \over U_p}}
e^{\sqrt{\tilde{\epsilon}_\perp U_p}s/k_{\rm B}T},
\label{WKBresult}
\end{equation}

Eq.\ref{WKBresult}\ may also be obtained from simple scaling
considerations.  The energy of an optimal kink is found by minimizing
\begin{equation}
f_1 \sim {\tilde{\epsilon}_\perp \over 2}\left({s \over w}\right)^2 w
+ U_p w
\label{kinkestimate}
\end{equation}
over the width (in the $y$ direction) $w$, giving $w^* \sim
\sqrt{\tilde{\epsilon}_\perp/U_p} s$ and $f_1^* \sim
\sqrt{\tilde{\epsilon}_\perp U_p}s$.  Such a kink occurs with a
probability proportional to $\exp(-f_1^*/k_{\rm B}T)$ in a length $w$.
The condition $L_{\rm kink}/w \exp(-f_1^*/k_{\rm B}T) \sim O(1)$ then
gives Eq.\ref{WKBresult}.

When the sample is
larger than $L_{\rm kink}$ along the field axis, the flux line will
wander as a function of $y$, with
\begin{equation}
\langle [z(y)-z(0)]^2 \rangle \sim D_zy,
\label{vortexdiffusion}
\end{equation}
where the ``diffusion constant'' $D_z \approx s^2/L_{\rm kink}$.

For $\sqrt{\tilde{\epsilon}_\perp U_p}s \lesssim k_{\rm B}T$, the
pinning is extremely weak, and the WKB approximation is no longer
valid.  Instead, the diffusion constant $D_z \approx k_{\rm
B}T/\tilde{\epsilon}_\perp$, as obtained from
Eq.\ref{singlevortexfreeenergy}\ with $U_p = 0$.  At much lower
temperatures, when $\xi_c \ll s$, the energy in Eq.\ref{upinning}\
must be replaced by the cost of creating a ``pancake''
vortex\cite{Clem}\ between the CuO$_2$ planes.  In this regime, $L_{\rm
kink} \sim \xi_{ab}(s/\xi_c)^{\epsilon_0 s /k_{\rm B}T}$.

For $T \approx 90K$, as in the experiments of Kwok et. al.\cite{Kwok}\,
$\xi_c/s \approx 2.3$, and Eq.\ref{upinning}\ gives
$\sqrt{\tilde{\epsilon}_\perp U_p}s/k_{\rm B}T \ll 1$, indicative of
weakly pinned vortices in the liquid state.  The
transverse wandering in this {\sl anisotropic} liquid is described by
a boson ``wavefunction'' with support over an elliptical region of
area $k_{\rm B}T
L_y/\sqrt{\tilde{\epsilon}_\parallel\tilde{\epsilon}_\perp}$ with
aspect ratio $\Delta x/\Delta z = \gamma^{-1} \approx 5$ for ${\rm
YBa_2Cu_3O_7}$.  For $L_y \approx 1mm$, a typical sample dimension along ${\bf
\hat{y}}$, the dimensions of this ellipse are of order microns.  Since
typical vortex spacings at the fields used in Ref.\onlinecite{Kwok}\
are of order $400\AA$, these flux lines are highly entangled.

To understand the bulk propertices of the vortex liquid, it is useful
to employ the hydrodynamic description of section \ref{hydromodel}.
In the liquid, the appropriate form of the free energy
is\cite{MNhydro,derivationnote}
\begin{eqnarray}
F_{\rm L} & = & {1 \over {2n_0^2}} \int {{d^3{\bf q}} \over {(2\pi)^3}} \bigg\{
c_{11}({\bf q})|\delta
n({\bf q})|^2 + c_{44,\parallel}({\bf q}) |\tau_x({\bf q})|^2
\nonumber \\
& & + c_{44,\perp}({\bf q}) |\tau_z({\bf q})|^2 \bigg\} - \int \!
d^3{\bf r} V_{\rm P}[z]  \delta n({\bf r}),
\label{liquidhydro}
\end{eqnarray}
where $\delta n = n - n_0$, with $n_0 = B_y/\phi_0$ the mean density.
Here the compression modulus $c_{11}$ and tilt moduli $c_{44,\perp}$ and
$c_{44,\parallel}$ are regular functions of ${\bf q}$ with finite
values at ${\bf q = 0}$.

On physical grounds, we expect intrinsic pinning to enter
Eq.\ref{liquidhydro}\ both through an increase in $c_{44,\perp}$,
which decreases fluctuations perpendicular to the layers, and through the
$\tilde{V}_{\rm P}$ term which tends to localize the vortices near the
minima in the periodic potential.

The former effect only acts to increase the anisotropy of the liquid.
The latter term, however, explicitly breaks translational symmetry
along the $z$ axis, inducing a modulation of the vortex density,
\begin{equation}
\int dz e^{-iq_z z} \langle \delta n({\bf r}) \rangle = {n_0^2 \over
{c_{11}(q_z,q_x=q_y=0)}} V_{\rm P}[q_z].
\label{densitymodulation}
\end{equation}
This modulation corrects the static structure function, $S({\bf q}) =
\langle \delta n({\bf q}) \delta n(-{\bf
q})\rangle/(2\pi)^3\delta^{(3)}({\bf q = 0})$, according to
\begin{equation}
S({\bf q}) = S_0({\bf q}) + {n_0^4 \over {[c_{11}(q_z)]^2}} |V_{\rm
P}[q_z]|^2 (2\pi)^2\delta(q_x)\delta(q_y),
\label{ssf}
\end{equation}
where $S_0(q)$ is the static structure function for $V_{\rm P}=0$.
Since $V_{\rm P}[z]$ is a periodic function, the correction term shows
peaks at the discrete reciprocal lattice vectors for which $q_z$ is an
integral multiple of $2\pi/s{\bf\hat{z}}$.

The situation is somewhat analogous to applying a weak uniform
field to a paramagnet, inducing a proportionate magnetization. Unlike the
magnetic case, however, the layering perturbation leaves a residual
translational symmetry under shifts $z \rightarrow z + s$.  It
is the breaking of this discrete group which we will identify with the
freezing of the vortex liquid.


\section{The Crystal Phase}
\label{CrystalPhaseSection}

Considerable work already exists on intrinsic
pinning in vortex crystals.\cite{intrinsicrefs}\ We review the essential
ideas here, and discuss its implications for thermal
fluctuations at low temperatures.

\subsection{Zero temperature properties}

To study the effect of layering upon the vortex state, we first consider
the limit of a weak periodic modulation of the order parameter along the
$z$--axis.  In this case, the resulting (zero temperature) configuration is
only slightly perturbed from the ideal lattice predicted by
Ginzburg--Landau (or London) theory.  The free energy in this case may be
written in terms of the phonon coordinates ${\bf u}({\bf r})$, as
\begin{equation}
F_{\rm elastic} =  \int {{d^3{\bf q}} \over {(2\pi)^3}} {K_{\alpha\beta
i j} \over 2}q_\alpha q_\beta u_i({\bf q})u_j(-{\bf q})  + F_{\rm IP},
\label{elasticenergy}
\end{equation}
where $i$ and $j = x,y$, $\alpha$ and $\beta = x,y,z$, and the
contribution to the free energy of the layering potential is
\begin{equation}
F_{\rm IP} = - \int \! dy \sum_{x_n,z_n} V_{\rm P}[z_n + u_z(x_n,y,z_n)].
\label{FIP}
\end{equation}
In general the elasticity theory is quite complex due to the
anisotropy and wavevector dependence on the scale of $\lambda$.
Rather than work with a specific form of the elastic moduli, we will
obtain general expressions in terms of an unspecified set of $K_{ij}({\bf
q}) \equiv K_{\alpha\beta i j}q_\alpha q_\beta$.

To obtain the correct continuum limit of Eq.\ref{FIP}, we consider the
possible commensurate states of the layers and vortex array.  The
triangular equilibrium lattice in this orientation is described by
the two lattice vectors ${\bf a}_1 =
C\gamma^{-1}\hat{\bf z}$, ${\bf a}_2 = C\gamma^{-1}/2(\hat{\bf z} +
\sqrt{3}\gamma^2\hat{\bf x})$, with $C^2 = 2\phi_0/\sqrt{3}B_y$.
Commensurability effects occur when the minimum $z$--displacement
between vortices, $C/(2\gamma) = (n/m)s$, where $n$ and $m$ are
integers, chosen relatively prime for definiteness.  This gives the
commensurate fields
\begin{equation}
B_y^{(m,n)} = {m \over n} {\phi_0 \over {2\sqrt{3}\gamma^2s^2}}.
\label{commensuratefields}
\end{equation}
For simplicity we
consider here only the integral states with $m=1$.  In
this case, $V_{\rm P}[z_n + u] = V_{\rm P}[u]$, and we can
straightforwardly take the continuum limit
\begin{equation}
F_{\rm IP} = - \int \! d^3{\bf r} V_{\rm P}[u_z({\bf r})].
\label{FIPcontinuum}
\end{equation}

Such an expression may be derived explicitly from the Abrikosov solution
for fields near $H_{c2}$,\cite{intrinsicrefs}\ in which case the
pinning potential is
\begin{equation}
V_{\rm P}[u] \approx {{Hj_c s} \over {2\pi c}}\cos(2\pi u_z/s),
\label{Ivlevapprox}
\end{equation}
where
\begin{equation}
j_c = {4 \over {\beta_{\rm A}\sqrt{\pi}}} {{c(H_{c2}-H)} \over
{\kappa^2 s}} {{\xi_c\gamma^2} \over s} \exp (-8\xi_c^2/s^2).
\label{jcdefinition}
\end{equation}
Here $\kappa = \lambda_{ab}/\xi_{ab}$ is the usual Ginzburg--Landau
parameter, and $\beta_{\rm A} \approx 1.16$.

{}From Eqs.\ref{elasticenergy}\ and \ref{FIPcontinuum}, it is clear that
for these commensurate fields, the ground state is unchanged, i.e.
${\bf u = 0}$.  Away from these fields, however, the fate of the
lattice is less obvious.  Ivlev et.  al.\cite{intrinsicrefs}\ have
shown that, for a small deviation from a commensurate field, it is
energetically favorable for the vortex lattice to shear in order to
remain commensurate with the copper--oxide plane spacing.  Because
such a distortion requires some additional free energy, it will
generally be favorable, in addition, for the internal magnetic
induction ${\bf B}$ to deviate from the applied field ${\bf H}$ to
allow a better fit to the crystal.  This Meissner--like effect will be
discussed in more detail in section \ref{ICphases}.

For strong layering, such as that described by the Lawrence--Doniach model,
the pinning effects are much more pronounced.  When $\xi_c \ll s$, the
magnetic field remains essentially confined between the $CuO_2$ layers, and
the vortex array is thus {\sl automatically} commensurate at all applied
fields.  Although such strong confinement of vortices can lead to interesting
non--equilibrium states,\cite{Levitov}\ we will confine our discussion to
equilibrium.

\subsection{Thermal Fluctuations about the commensurate state}

Thermal fluctuations of the vortex lattice are described by the partition
function
\begin{equation}
{\cal Z}_{\rm elastic} = \int [du({\bf r})] \exp\left( -F_{\rm
elastic}/k_{\rm B}T \right).
\label{elasticpartitionfunction}
\end{equation}

In three dimensions, phonon fluctuations are small, and expanding the
pinning potential around its minimum for small ${\bf u}$, gives the
quadratic free energy
\begin{equation}
F_{\rm elastic} \approx \int {{d^3{\bf q}} \over {(2\pi)^3}} {K_{i
j}({\bf q}) \over 2}u_i({\bf q})u_j(-{\bf q})  +
{\Delta \over 2} |u_z({\bf q})|^2,
\label{quadraticefe}
\end{equation}
where $\Delta \equiv -V_{\rm P}^{''}[z=0]$.

The displacement field fluctuations can be calculated from
Eq.\ref{quadraticefe}\ by equipartition, yielding the general result
\begin{eqnarray}
{{\langle u_x^2 \rangle} \over {k_{\rm B}T}}  & = & \int_{\rm BZ} \! {{d^3{\bf
q}}
\over (2\pi)^3}
{{K_{zz}({\bf q}) + \Delta} \over {K_{xx}({\bf q})(K_{zz}({\bf
q})+\Delta) - [K_{xz}({\bf q})]^2}}, \label{uxfluctuations}\\
{{\langle u_z^2 \rangle} \over {k_{\rm B}T}} & = & \int_{\rm BZ} \! {{d^3{\bf
q}}
\over (2\pi)^3}
{{K_{xx}({\bf q})} \over {K_{xx}({\bf q})(K_{zz}({\bf q})+\Delta) -
[K_{xz}({\bf q})]^2}}, \label{uzfluctuations}
\end{eqnarray}
where BZ indicates an integral over the Brillouin zone.

The effect of the periodic potential is thus to uniformly decrease the
fluctuations of $u_z$ at all wavevectors.  For $\Delta
\gtrsim B^2/\lambda^2$, this decrease is substantial over the entire
Brillouin zone, and
\begin{eqnarray}
\langle u_x^2 \rangle & \approx &  k_{\rm B}T \int_{\rm BZ} \! {{d^3{\bf q}}
\over (2\pi)^3} {1 \over {K_{xx}({\bf q})}} , \label{uxf1}\\
\langle u_z^2 \rangle & \approx &  k_{\rm B}T \int_{\rm BZ} \! {{d^3{\bf q}}
\over (2\pi)^3} {1 \over \Delta}. \label{uzf1}
\end{eqnarray}
In stronger fields, for $\Delta \lesssim
B^2/\lambda^2$, the only the contributions from $q \lesssim
\sqrt{\Delta} /B$ are strongly suppressed.  For $1-B/H_{c2} \ll 1$,
Eqs.\ref{Ivlevapprox}\ and \ref{jcdefinition}\ can be combined to give
the ratio
\begin{equation}
\Delta\lambda^2/B^2 \approx {2 \over {\pi^{3/2}\beta_{\rm A}}}
(H_{c2}/B - 1)\left({\xi_c \over s}\right)^3 e^{-8(\xi_c/s)^2}.
\label{characteristicratio}
\end{equation}
At lower fields and temperatures, one expects the mean--field estimate
above to break down and $\Delta\lambda^2/B^2$ to increase, possibly
settling down to a constant value at low temperatures.

For magnetic fields oriented along the $c$ axis, the Lindemann criterion
has been used to estimate the melting point of the vortex
lattice\cite{NelsonSeung}\ by requiring that $\langle |{\bf u_i}|^2 \rangle
= c_{\rm L}^2 a_i^2$, for $i=x,z$, with a ``Lindemann number'' $c_{\rm L}
\approx 0.2-0.4$.  As is clear from
Eqs.\ref{uxfluctuations}--\ref{uzfluctuations}, once layering is
included, the increased stiffness for $u_z$ makes the two ratios
$\langle u_x^2 \rangle/a_x^2$ and $\langle u_z^2 \rangle/a_z^2$
unequal.  Indeed, the second ratio is strongly suppressed relative to
the first.  Extending the Lindemann criterion to this situation
suggests that the strains in $u_x$ might be alleviated by a partial
melting of the lattice without affecting the broken symmetry leading
to the $u_z$ displacements.  Such
a scenario corresponds to the unbinding of dislocations with Burger's
vectors along the $x$ axis.  The phase in which these dislocations are
unbound is the smectic.

\subsection{Strongly Layered Limit}
\label{stronglayeringsection}

To further elucidate the nature of the smectic phase, it is helpful to
discuss the limit of very strong layering.  In this
case, the vortex lines are almost completely confined within the spaces
between neighboring $CuO_2$ layers.  For moderate fields, occupied
layers will be separated by separated by several unoccupied ones, and the
interactions between vortices in different layers may be considered
weak.  Because of the strong layering, the out--of--plane component of
the displacement field $u_z$ is suppressed, so that the free
energy of the system may be written to a first approximation as
\begin{equation}
F_{\rm layers} = \sum_n {1 \over 2}\int {{d^2{\bf q_{\perp}}} \over {(2\pi)^2}}
\left(K_x q_x^2+K_z q_z^2\right)|u_n({\bf
q}_\perp)|^2,
\label{layerfreeenergy}
\end{equation}
where ${\bf q}_\perp = (q_x,q_y)$ and $u_n({\bf q}_\perp) = u_x({\bf
q}_\perp, ns)$.  For a qualitative discussion of smectic ordering, it
is sufficient to take $K_x$ and $K_y$ independent of ${\bf q}$.

Eq.\ref{layerfreeenergy}\ neglects both inter--layer interactions and
hopping.  The former are included perturbatively via the free energy
\begin{equation}
F_{\rm int.} = - \sum_n \int \! dx dy
v_{\rm IL}
\cos {{2\pi} \over a}(u_{n+1} - u_n),
\label{interlayerinteractions}
\end{equation}
where $v_{\rm IL}$ is an inter--layer interaction energy and $a$ is
the lattice spacing in the $x$ direction.  The periodic form of
the interaction is required by the symmetry under lattice translations
$u \rightarrow u + a$ within each layer.

Once hopping of flux lines between neighboring occupied layers is
included, $u_n$ is no longer single valued within a given layer.  In
fact, a configuration in which a single line hops from layer $n$ to
layer $n+1$ corresponds to a dislocation in layer $n$ paired with an
anti--dislocation in layer $n+1$, since
\begin{equation}
\oint \bbox{\nabla} u_k \cdot d\bbox{\ell} = a(\delta_{k,n} - \delta_{k,n+1})
\label{burgerscircuit}
\end{equation}
for a contour surrounding the hopping point (see
Fig.\ref{hoppingfigure}).  Such dislocation--antidislocation pairs,
which we will refer to as large kinks, can be created in neighboring
layers with a dislocation fugacity $y_d =
\exp(-E_{\rm lk}/k_{\rm B}T)$, where the core energy
\begin{equation}
E_{\rm lk} \approx \sqrt{\tilde{\epsilon}_\perp U_p}ms,
\label{lkestimate}
\end{equation}
as estimated from Eq.\ref{kinkestimate}\ with $s \rightarrow ms$.  Note
that the dislocation and anti--dislocation must have the same $x$ and
$y$ coordinates, since misalignment is accompanied by an energy cost
proportional to the extra length of vortex between the occupied
layers.

The full theory described by Eqs.\ref{layerfreeenergy}\ and
\ref{burgerscircuit}\ plus dislocations can be studied using
a perturbative renormalization group (RG) expansion in $v_{\rm IL}$
and $y_d$, using techniques developed for the XY model in a
symmetry--breaking field.\cite{JKKN}\ For $y_d=v_{\rm IL}=0$, the
Gaussian free energy of Eq.\ref{layerfreeenergy}\ describes a fixed
line of independently fluctuating vortex layers parameterized by the
dimensionless ratio $\sqrt{K_xK_y}/k_{\rm B}T$.  To characterize the
order along this fixed line, we define a translational order parameter
(characterizing correlations along the $x$ axis)
within the $n^{\rm th}$ layer by summing over vortex lines according
to
\begin{equation}
\rho_\parallel(x,y,n) \equiv \sum_k \exp(2\pi i x_k^{(n)}(y)/a),
\label{inplaneorder}
\end{equation}
where $x_k^{(n)}(y)$ is the coordinate of the $k^{\rm th}$ vortex line
in layer $n$ at a length $y$ along the field direction.
The correlation function $C_{\rm T,\parallel}(x,y,n) \equiv \langle
\rho_\parallel(x,y,n)\rho_\parallel(0,0,0)\rangle$ is then evaluated by
inserting $x_k^{(n)}(y)
= ka + u_n(ka,y)$ and converting the sum to an integral via $\sum_k
\rightarrow \int dx/a$.  One finds
\begin{equation}
C_{\rm T,\parallel}(x,y,n) \sim \left( {K \over {K_x x^2+K_y y^2}}\right)^{-
\pi k_{\rm B}T/Ka^2}\delta_{n,0},
\label{QLRO}
\end{equation}
i.e. quasi--long--range order within the planes.

This fixed line is {\sl always} unstable either to interlayer
couplings, to dislocations, or to both perturbations.  The linear (in
$y$ and $v_{\rm IL}$) RG flows which determine the stability are
\begin{eqnarray}
{{dv_{\rm IL}} \over {dl}} & = & \left(2 - {{2\pi k_{\rm B}T} \over
{Ka^2}}\right)v_{\rm IL}, \label{vILrecursion}\\
{{dy_d} \over {dl}} & = & \left(2 - {{Ka^2} \over {2\pi k_{\rm
B}T}}\right)y_d, \label{yrecursion}
\end{eqnarray}
where $K \equiv \sqrt{K_xK_y}$, and $l = \ln(b/a)$ is the logarithm of
the coarse--graining length scale $b$.

When $k_{\rm B}T < Ka^2/4\pi$, dislocations are irrelevant at the fixed
line, so that $y$ decreases under renormalization.  In this regime,
however, $v_{\rm IL}$ increases with $l$, so that interactions between
the layers are important for the large distance physics.  To study
this regime, one may therefore expand the cosine of
Eq.\ref{interlayerinteractions}\ in $u_{n+1} -u_n$, obtaining a
discrete version the usual three--dimensional elastic theory.  In this
limit,
\begin{equation}
C_{\rm T,\parallel}(x,y,n) \sim {\rm Const.}
\label{LROlayered}
\end{equation}
for large $|x|$, $|y|$, or $|n|$.

At high temperatures, when $k_{\rm B}T > Ka^2/\pi$, $v_{\rm IL}$ scales to
zero while the fugacity $y_d$ is relevant.  Unbound dislocations on long
length scales therefore invalidate the elastic theory of
Eq.\ref{layerfreeenergy}.  Following standard arguments, the translational
correlation function in such an unbound vortex plasma becomes exponentially
small, i.e.
\begin{equation} C_{\rm T,\parallel}(x,y,n) \sim
\exp(-\tilde{r}/\xi_{\rm T}),
\label{liquidlayered}
\end{equation}
where $\tilde{r} \equiv \sqrt{ (K_x/K) x^2 + (K_y/K) y^2 + \chi
s^2n^2}$ with $\chi$ a constant, and $\xi_{\rm T}$ is a finite
translational correlation length.

For temperatures in the intermediate range $Ka^2/4\pi < k_{\rm B}T <
Ka^2/\pi$, {\sl both} $y_d$ and $v_{\rm IL}$ are relevant operators.  The
eventual nature of the ordering at long distances presumably takes one of
the two above forms, though the critical boundary at which the system
loses long--range translational order in the $x$ direction is not
accessible by this method.

It remains to discuss translational order along the
layering axis.  Such order is characterized by the parameter
\begin{equation}
\rho_\perp(x,y,n) \equiv \sum_k \exp(2\pi i z_k^{(n)}(y)/a_z),
\label{transverseordering}
\end{equation}
where $a_z$ is the distance between occupied vortex layers (and therefore
an integral multiple of the CuO$_2$ plane spacing, i.e.  $a_z = ms$).
Because we have, by construction, confined the vortices to these layers,
however, $z_k^{(n)} = na_z$ for every $k$, and the exponential in
Eq.\ref{transverseordering}\ is always unity.  Both phases described above,
regardless of the relevance of $v_{\rm IL}$ and $y$, therefore retain
long--range order along the $z$ axis.

Within the strongly layered model, there are still excitations which can
destroy this transverse ordering.  These are configurations in which a flux
line hops out of an occupied layer into one of the ($m-1$) unoccupied
intermediate planes between it and the next occupied layer.  Such an
excursion costs an energy proportional to the length of the vortex segment
in the unoccupied layer, so that only short intermediate segments occur at low
temperatures.  These out--of--plane hops reduce the amplitude $\langle
\rho_\perp \rangle$, but do not drive it to zero.  At very high
temperatures, entropy may counterbalance this energy and drive the free
energy cost for such intervening vortices negative.  Once this occurs,
translational order will be lost along the $z$ axis as well, and the system
will be a true liquid.  Nevertheless, at intermediate temperatures above
the unbinding transition for dislocations in the $u$ field but below the
temperature at which infinite vortices enter the intermediate copper oxide
planes, we expect the system to sustain ``one--dimensional'' long--range
order along the $z$ axis, i.e. a smectic state.

\begin{figure}
\caption{Hopping of a vortex line between neighboring vortex layers.}
\label{hoppingfigure}
\end{figure}

\section{Critical Behavior}
\label{criticalsection}

Having established the possibility of a smectic phase approaching both
from the crystalline and liquid limits, we now focus on the critical
behavior near the putative liquid--smectic transition, using a Landau
order parameter theory.  A closely related Landau theory which
describes a low temperature smectic--crystal transition is discussed
in Appendix \ref{sxappendix}.  The natural order
parameter to describe the smectic ordering is $\rho_\perp$ defined in
Eq.\ref{transverseordering}.  To simplify notation, we define a new field
$\Phi = \rho_\perp$, so that, in the continuum notation (i.e. outside the
strongly layered limit),
\begin{equation}
n ({\bf r}) \approx n_0 {\rm Re} \left\{ 1 + \Phi({\bf r})e^{-iqz}
\right\},
\label{densitywave}
\end{equation}
where $n_0$ is the background density, and $q = 2\pi/a_z$ is the wavevector
of the smectic layering.  The complex translational order parameter
$\Phi({\bf r})$ is assumed to vary slowly in space.  The superconductor is
invariant under translations and inversions in $x$ and $y$, and has a
discrete translational symmetry under $z \rightarrow z + s$, where $s$ is
the CuO$_2$ double--layer spacing.  From Eq.\ref{densitywave}, these
periodic translations correspond to the phase shifts $\Phi \rightarrow \Phi
e^{-iqs}$.  We continue to assume, as in the previous section, that $a_z=ms$,
with $m$ an arbitrary integer.  The most general free energy consistent
with these symmetries is
\begin{eqnarray}
F & = \int \! d^3\!{\bf r} \left\{ \right. & {K \over 2}
|(\bbox{\nabla}-i{\bf A})
\Phi|^2 + {r \over 2}|\Phi|^2 + {v \over 4}|\Phi|^4 \nonumber\\
& & \left. - {g \over
2}\left( \Phi^m + \Phi^{*m} \right) + \cdots \right\},
\label{criticalfreeenergy}
\end{eqnarray}
where the coordinates have been rescaled to obtain an isotropic
gradient term.  The ``vector potential'' ${\bf A}$ represents changes
in the applied field $\delta {\bf H} = \delta H_b {\bf\hat{y}} + H_c
{\bf\hat{z}}$, with $A_x = 0$, $A_y = q H_c/H_b$, and $A_z = q\delta
H_b/H_b$.  The form of this coupling follows from the transformation
properties of $\Phi$.\cite{DeGennes,Transformnote}\

Eq.\ref{criticalfreeenergy}\ assumes a {\sl local} form of the free energy.
Additional non--local interactions arise due to interactions with long
wavelength fluctuations in the density and tangent fields.  The most
relevant (near the critical point) of these couplings is
\begin{equation}
F_{\Phi-\delta n} = -\gamma \int \! d^3\!{\bf r} \delta n |\Phi|^2,
\label{pdncoupling}
\end{equation}
where the correlations of $\delta n$ are determined from
Eqs.\ref{liquidhydro}\ and \ref{constraint}.

When $\delta {\bf H} = {\bf A} = \gamma = 0$, Eq.\ref{criticalfreeenergy}\
is the free energy of an XY model with an $m$--fold symmetry breaking
term.
A second order freezing transition occurs within Landau theory when
$v>0$ and $r \propto T-T_s$ changes sign from positive (in the liquid)
to negative (in the smectic).  The renormalization group (RG) scaling
dimension, $\lambda_m$, of the symmetry breaking term is known {\sl
experimentally} in three dimensions to be $\lambda_m \approx 3 - 0.515m -
0.152m(m-1)$.\cite{AA}\  For $m>m_c \approx 3.41$, the field $g$ is
irrelevant ($\lambda_m <0$), and the transition is in the XY
universality class.\cite{smallmfoot}\  The magnetic fields used by Kwok
et. al.\cite{Kwok}\ correspond to $m = 9-11$,\cite{AGL}\ well into this
regime.  The static critical behavior is characterized by the
correlation length exponent $\nu \approx 0.671 \pm 0.005$ and
algebraic decay of order parameter correlations at $T_s$,
\begin{equation}
\left\langle \Phi({\bf r})\Phi^*({\bf 0}) \right\rangle \sim {1 \over
r^{1+\eta}},
\label{criticalcorrelations}
\end{equation}
with $\eta \approx 0.040 \pm 0.003$.\cite{exponentrefs}\

To study the effects of coupling to long wavelength fluctuations when
$\gamma \neq 0$, we first satisfy Eq.\ref{constraint}\ by defining an
auxiliary ``displacement''--like field ${\bf w}$ via
\begin{eqnarray}
\delta n & = & - \bbox{\nabla}_\perp \cdot {\bf w}  \nonumber \\
\bbox{\tau} & = & \partial_y {\bf w}.
\label{wdef}
\end{eqnarray}
After this change of variables, Eq.\ref{liquidhydro}\ becomes
\begin{eqnarray}
F_{\bf w} & = & {1 \over {2n_0^2}} \int d^3{\bf r} \bigg\{
c_{11}|\bbox{\nabla}_\perp
\cdot {\bf w}|^2 + c_{44,\parallel}|\partial_y w_x|^2
\nonumber \\
& & + c_{44,\perp}|\partial_y w_z|^2 \bigg\},
\label{wfreeenergy}
\end{eqnarray}
where we have taken the ${\bf q}=0$ limits of the elastic moduli to study
the critical behavior, and dropped the $V_{\rm IP}$ term which only
couples to ${\bf w}$ at finite $q_z$.  Eq.\ref{pdncoupling}\ then becomes
\begin{equation}
F_{\Phi-{\bf w}} = \gamma\int d^3{\bf r} \bbox{\nabla}_\perp \cdot {\bf w}
|\Phi|^2.
\label{pwcoupling}
\end{equation}

Eq.\ref{pwcoupling}\ is an anisotropic form of a coupling studied
previously in the context of the compressible Ising model, in which ${\bf
w}$ describes the phonon modes of a compressible lattice on which the spins
reside.\cite{CIsing}\ As shown in appendix \ref{murelevanceappendix}, the
techniques developed for that problem give the renormalization group
eigenvalue $\lambda_\gamma = \alpha/2\nu$ for this coupling at the critical
point.  Since $\alpha = 2-3\nu \approx -0.01$ is negative, the long
wavelength density fluctuations are irrelevant for the critical behavior.

\section{Smectic Phase}
\label{smecticsection}

\subsection{Static Behavior}
\label{staticbehaviorsection}

Deep in the ordered phase ($r<0$), amplitude fluctuations of $\Phi$
are frozen out.  Writing $\Phi = \sqrt{|r|/v} e^{2\pi i u/a}$,
Eq.\ref{criticalfreeenergy}\ becomes, up to an additive constant,
\begin{equation}
F_{\rm smectic} = \int \! d^3\!{\bf r} \left\{ {\kappa \over 2}
(\bbox{\nabla}u \!-\! {\bf\cal A})^2 - \tilde{g}\cos 2\pi u/s \right\},
\label{smecticfreeenergy}
\end{equation}
where $a=ms$, $\kappa = 4\pi^2|r|K/a^2v$, $\tilde{g} = g(|r|/v)^{m/2}$, and
the reduced vector potential is ${\bf\cal A} = {\bf A}/q$.  The
displacement field $u$ describes the deviations of the smectic layers
from their uniform state.  The sine--Gordon term is an effective
periodic potential acting on these layers.  As is well known from the
study of the roughening transition,\cite{roughening}\ such a
perturbation is always relevant in three dimensions.  The smectic
state is thus {\sl pinned} at long distances (i.e. the displacements
$u$ of each smectic layer are localized in a single minima of the
cosine).

\begin{figure}
\caption{Fluctuating density wave in the smectic state.  Displacements
of the layers from their mean positions are described by the field $u$.}
\label{layerfig}\end{figure}

To further characterize the smectic phase, we consider the transverse
magnetic susceptibility, which defines the macroscopic tilt modulus,
\begin{equation}
c_{44, \perp}^{-1} \equiv \left.{{\partial B_c} \over {\partial
H_c}}\right|_{H_c = 0}.
\label{c44def}
\end{equation}
The field $H_c$ attempts to tilt the smectic layers.  However,
${\bf\cal A} \propto H_c$ is an {\sl irrelevant} operator in the
smectic phase (as can be easily seen by replacing the periodic
potential by a ``mass'' term $\propto u^2$).  This implies that the
smectic layers do not tilt under weak applied fields, i.e. $\partial
\langle \partial_y u \rangle/\partial H_c|_{H_c =0} = 0$.  Naively,
this implies an infinite tilt modulus.

A more careful treatment shows that $c_{44, \perp}$ actually remains
finite in the smectic phase.  To compute $c_{44, \perp}$ from first
principles, we use the thermodynamic relation
\begin{equation}
B_c = -4\pi {{\partial f} \over {\partial H_c}},
\label{Bcthermo}
\end{equation}
where $f$ is the {\sl full} free energy of the system, including
a smooth part $f_0$ not involving $\Phi$ and not included in
Eq.\ref{criticalfreeenergy}, i.e.
\begin{equation}
f = f_0 - k_{\rm B}T \ln {\cal Z}_\Phi.
\label{ftotaldef}
\end{equation}
To evaluate Eq.\ref{Bcthermo}, we need to consider in detail the
dependence of the free energy {\sl and} the coefficients in
Eq.\ref{criticalfreeenergy}\ on $H_c$.  This dependence arises in two
ways, because an applied $H_c$ can be decomposed into a rotation and a
scaling of the full field ${\bf H}$.  If the system were fully
rotational invariant, the rotational part would enter $F$ purely
through the ``gauge--invariant'' coupling to ${\bf \cal A}$ of
Eq.\ref{criticalfreeenergy}.  However, anisotropy breaks this
invariance, leaving instead only an inversion symmetry under $z
\rightarrow -z$.  The inversion symmetry allows for a quadratic
dependence of $r$ and of $f_0$ on $H_c$.\cite{hdepnote}\ The scaling
part also contributes quadratic dependence, which may be combined with
the previous effect.  Taking both into account, and matching the tilt
modulus to the tilt modulus of the liquid phase (i.e. with $\Phi = 0$)
leads to
\begin{equation}
B_c = {{Kq} \over H_b}{\rm Im}\left(\Phi^*\partial_y\Phi\right) +
\left(c_{44,\perp 0}^{-1} - r'' |\Phi|^2\right)H_c,
\label{transversefield}
\end{equation}
where $c_{44,\perp 0}$ is the tilt modulus obtained from anisotropic GL
theory (without accounting for the discreteness of the layers) and
$r'' \equiv \partial^2 r/\partial H_c^2|_{H_c = 0}$.

Eq.\ref{transversefield}\ has a simple physical interpretation.  The
first term is the contribution to $B_c$ from tilting of the layers
(described by a phase shift of $\Phi$).  This term is zero for small
fields $H_c$ due to the cosine pinning potential.  Even when the
layers retain a fixed orientation perpendicular to the $c$ axis,
however, the transverse field can penetrate via the second term.  Such
motion arises microscopically from a non--zero equilibrium
concentration of vortices with large kinks extending between
neighboring smectic layers, as suggested in section
\ref{stronglayeringsection}.  Eq.\ref{transversefield}\ predicts a
non--divergent singularity $c_{44,\perp}(T) - c_{44,\perp}(T_s) \sim
|T-T_s|^{1-\alpha}$ at the critical point, where $\alpha$ is the
specific heat exponent.

At low temperatures in the smectic phase, we can estimate the tilt
modulus in terms of properties of kinks.  In zero field, the concentrations
of large kinks carrying magnetic field in the $+{\bf \hat{z}}$ and
$-{\bf\hat{z}}$ directions are equal, leading to zero net field along
the $c$ direction.  For $H_c \neq 0$, the energy of a kink depends
upon its orientation due to the $-B_c H_c/4\pi$ term in the GL free
energy, yielding
\begin{equation}
E_{\pm} \approx E_{\rm lk} \pm ms\phi_0 H_c/4\pi.
\label{epm}
\end{equation}
The difference in the concentrations of up and down kinks takes the
activated form
\begin{equation}
n_+ - n_- \sim {B \over {\phi_0 w_{\rm lk}}} e^{-E_{\rm lk}/k_{\rm B}T} \sinh
\left( {{ms\phi_0 H_c} \over {4\pi k_{\rm B}T}} \right),
\label{kinkdiff}
\end{equation}
where $w_{\rm lk} \sim \sqrt{\tilde{\epsilon}_\perp/U_p}ms$, estimated
from Eq.\ref{kinkestimate}\ with $s \rightarrow ms$.  Since $B_c =
(n_+ - n_-)ms\phi_0$, Eq.\ref{c44def}\ yields
\begin{equation}
c_{44,\perp} \approx \left[{{\sqrt{\tilde{\epsilon}_\perp/U_p}k_{\rm
B}T} \over {B\phi_0 ms}}\right]e^{E_{\rm lk}/k_{\rm B}T},
\label{c44perpkinks}
\end{equation}
i.e. a large but finite tilt modulus.

\subsection{Dynamical Behavior}
\label{dynamicbehaviorsection}

Very similar phenomena occur in the dynamics of the smectic phase.
To study them, we need the equation of motion for $\Phi$.  On the
basis of symmetry and the lack of obvious conservation laws, a natural
conjecture is that of overdamped ``model A''\cite{HalperinHohenberg}\
dynamics.  Indeed, a careful treatment using
the general formalism of section \ref{hydromodel}\ gives (see appendix
\ref{eomappendix})
\begin{equation}
\gamma_{\rm BS}\partial_t \Phi = - 4q^2{{\delta F_{\rm crit.}}
\over {\delta\Phi^*}} + i\mu J_x \Phi - \tilde{\eta},
\label{criticaldynamics}
\end{equation}
where $\mu = q\phi_0 n_0/c$ and $\tilde{\eta}({\bf k}) = in_0 q
\eta_z(q{\bf\hat{z}} + {\bf k})$.  Eq.\ref{criticaldynamics}\
is remarkably similar to the model E dynamics\cite{HalperinHohenberg}\ for
the complex ``superfluid'' order parameter $\Phi$, where now $J_x$
plays the role of the ``electric field'' in the Josephson
coupling.\cite{conserveddensitynote}\ The actual electric field is
${\cal E}_x = j_{v,z}\phi_0/c$, leading via Eq.\ref{constitutive}\ to
(see appendix \ref{eomappendix})
\begin{eqnarray}
{\cal E}_x & \approx & -{{n_0\phi_0} \over
{2qc}}{\rm Im}\left(\Phi^* \partial_t\Phi\right)
\nonumber \\
& & + (1 -
|\Phi|^2/2)\left({B \over {H_{c2}}}\right)\rho_{xx,n}J_x,
\label{vortexcurrent}
\end{eqnarray}
where $\rho_{xx,n}$ is the normal state resistivity in the $x$
direction, whose appearance in the last term follows from the relation
$(n_0\phi_0/c)^2/\gamma_{\rm BS} \approx (B/H_{c2})\rho_{xx,n}$.

\begin{figure}
\caption{Sliding of a kink (thick curved line) along the field
direction, viewed along the $x$ axis for the case $m=2$.  Dashed lines
indicate the copper--oxide layers.  As the kink moves along the $y$
axis, net vorticity is transported in the $z$ direction.  Such motion
produces a finite resistivity in the smectic phase.}
\label{kinkslidefig}
\end{figure}

Eq.\ref{vortexcurrent}\ is interpreted in close analogy with
Eq.\ref{transversefield}.  In the absence of pinning due to the
periodic potential in Eq.\ref{smecticfreeenergy}, an applied force
induces a uniform translation of the layers, and thus a net transport
of vortices.  In the ordered phase, where $\Phi = \sqrt{|r|/v} e^{2\pi
i u/a}$, the first term in Eq.\ref{vortexcurrent}\ becomes
proportional to the velocity $\partial_t u$.  The second term
contributes even when the layers are constant.  It results
microscopically from the motion of equilibrium vortex kinks, which can
slide unimpeded along the $y$ axis and thereby transport vorticity
along the $z$ axis (see Fig.\ref{kinkslidefig}).  Such flow at
``constant structure'' is analogous to the permeation mode in smectic
liquid crystals.\cite{DeGennes}\

The presence of this defective motion implies a small but
non--zero resistivity at the L--S transition.  Near $T_s$,
Eq.\ref{vortexcurrent}\ predicts a singular decrease of the form
$\rho_{xx}(T) - \rho_{xx}(T_s) \sim |T_s - T|^{1-\alpha}$, similar to
the behavior of the tilt modulus.  At lower
temperatures (but still within the S phase) transport occurs via two
channels.  The permeation mode gives an exponentially small linear
resistivity $\rho_{xx} \sim \exp(-E_{\rm lk}/k_{\rm B}T)$ (above $T_s$,
single layer kinks give $\rho_{xx} \sim \exp(-E_{\rm k}/k_{\rm B}T)$,
with $E_{\rm k} \approx E_{\rm lk}/m$).

Non--linear transport occurs in parallel to the above linear
processes, via thermally activated liberation of vortex droplets,
inside which $u$ (or $u_z$ in the crystal phase) is shifted by $s$
(see Fig.\ref{dropletfig}).  Such a droplet costs a surface energy,
due to the creation of a domain wall between smectic regions shifted
by $s$.  The domain wall surface tension $\sigma_0$ is estimated from
\begin{equation}
\sigma_0 \sim {\kappa \over 2} \left( {s \over w} \right)^2 + \tilde{g} w,
\label{dwenergy}
\end{equation}
where $w$ is the width of the domain wall.  The first term represents
the elastic cost of the shift in $u$, while the second is the pinning
energy.  Minimizing Eq.\ref{dwenergy}\ gives $w \sim
\sqrt{\kappa/\tilde{g}}s$ and  $\sigma_0 \sim \sqrt{\kappa\tilde{g}}s$.
This surface energy must be balanced against the Lorentz force in the
interior, so that the energy of a droplet of linear size $L$ is
\begin{equation}
E_{\rm droplet} \sim \sigma_0 L^2 - {{JBs} \over c}L^3.
\label{dropletenergy}
\end{equation}

Eq.\ref{dropletenergy}\ gives a critical droplet size $L_c \sim
\sqrt{\kappa\gamma}c/(JB)$ and an energy barrier $E_{\rm B} \sim
(c/JB)^2(\kappa\gamma)^{3/2}s$.  Thermal activation therefore gives
\begin{equation}
{\cal E}_{nl} \sim e^{-(J_c/J)^2},
\label{activatedIV}
\end{equation}
where $J_c \sim (c/B)(\kappa\tilde{g})^{3/4}(s/k_{\rm B}T)^{1/2}$.  Similar
non--linear IV relations have been obtained previously for vortex/Bose
glasses,\cite{VGtheory,BGtheory}\ but our result is more closely
related to surface mobility below the roughening transition on crystal
surfaces.\cite{roughening}\  Unlike these proposed glass phases, the
smectic should always exhibit a nonzero linear resistivity as ${\bf
J}\rightarrow 0$.

\begin{figure}
\caption{A two dimensional cut through a droplet configuration of the
smectic layers, drawn for the case $m=2$.  The full three--dimensional
droplet has a spherical droplet.  Inside the droplet (outlined in
gray), the layers are shifted by one CuO$_2$ double layer spacing, $u
\rightarrow u + s$.}
\label{dropletfig}\end{figure}

\section{Supersolid Order and the Smectic to Crystal Transition}
\label{supersolidsection}

\subsection{Supersolid Nature of the Smectic Phase}

In sections \ref{staticbehaviorsection}\ and
\ref{dynamicbehaviorsection}, we have seen that the response functions
in the smectic phase retain many of the features of the vortex liquid.
Both the tilt modulus and conductivity remain finite, despite the
pinning of the smectic density wave by the CuO$_2$ layers.  As
discussed earlier, both phenomena are explained by the existence of an
equilibrium concentration of large vortex kinks extending between successive
occupied vortex layers.  These kinks facilitate both transverse
magnetic penetration and dissipation for currents along the $x$ axis.

This behavior is strikingly similar to the picture of ``supersolid''
vortex arrays recently proposed in Ref.\onlinecite{FNF}, for fields
parallel to the $c$--axis.  In the supersolid, a finite concentration
of interstitials or vortices are present in the vortex lattice, and
both the tilt modulus and conductivity in the presence of weak pinning
are finite.  Such a supersolid phase is distinct from the Abrikosov
solid in that it supportslong range crystalline order coexists with a
finite expectation value of the boson order parameter $\psi$, i.e.
\begin{equation}
\langle \psi({\bf r}) \psi^*({\bf 0}) \rangle \rightarrow {\rm
Const.}
\label{supersoliddef}
\end{equation}
as ${\bf r} \rightarrow \infty$.  The supersolid must occur at
sufficiently high magnetic fields, but its existence elsewhere in the
phase diagram seems unlikely.\cite{FNF}\

Using this characterization of broken $U(1)$ symmetry (under $\psi
\rightarrow \psi e^{i\theta}$), the vortex smectic is {\sl always} in
a supersolid phase.  As in Ref.\onlinecite{FNF}, this can be seen by
considering the correlation function of $\psi$'s.  Note that
$\psi({\bf r})$ destroys a vortex line at position ${\bf r}$ are
$\psi^*({\bf r})$ creates a line in the coherent state path integral
formalism.  Because there is always a finite probability of finding a
kink connecting the points ${\bf 0}$ and ${\bf r}$,
Eq.\ref{supersoliddef}\ is indeed satisfied.  With this understanding,
the second terms in Eqs.\ref{vortexcurrent}\ and
\ref{transversefield}\ have an additional complementary interpretation.
They correspond to the contributions from the ``superfluid fraction''
of a two--fluid system with ``superfluid'' (kink) and ``normal''
(smectic) parts.

In addition, the concept of symmetry breaking implies that a
continuous transition from the flux liquid state (with
$\langle\psi\rangle \neq 0$) to a smectic (translationally ordered in
one direction) phase must necessarily retain supersolid order.  For
the smectic phase to appear with $\langle\psi\rangle = 0$ would
require {\sl simultaneous} breaking of the discrete translation group
and restoration of the $U(1)$ symmetry.  Such a double critical point
can only occur by tuning two parameters (one in addition to the
temperature) or through a first order transition.  The physical
arguments of sections
\ref{stronglayeringsection} and
\ref{staticbehaviorsection}--\ref{dynamicbehaviorsection}, of course,
imply the stronger condition that the smectic phase must be supersolid
at {\sl all} temperatures.

\subsection{Consequences for Further Transitions at Low Temperatures}

At lower temperatures, provided point disorder remains negligible, the vortices
will order along the $x$ axis as well.\cite{Feinberg}\  What is the
nature of this two--dimensionally ordered phase?

The different possibilities may be classified by the order in which
the symmetries are broken.  At the lowest temperatures, we expect the
system to prefer a true solid phase, with broken translational order
in both directions (in particular $\langle\rho_\parallel\rangle \neq
0$, where $\rho_\parallel$ is the amplitude for periodic density
variations along the $x$--axis.  Recall that $\rho_\perp \equiv \Phi$
is the amplitude for density waves along $z$.) and a restored $U(1)$
symmetry (i.e. no interstitials).\cite{u1note}\ To connect this state
with the smectic phase in which $\langle \rho_\parallel \rangle = 0$
and $\langle \psi
\rangle \neq 0$ requires two changes of symmetry.

Three monotonic choices of symmetry breaking are shown in
Fig.\ref{sbfig}.  In the scenario (a), upon lowering the temperature
from the smectic phase first the $U(1)$ symmetry is restored, and
the translational symmetry along the $x$ axis is broken at a lower
temperature.  As remarked in the previous section, however, the
intermediate non--supersolid smectic phase that appears in this
sequence is impossible, so this sequence cannot occur.

Two physical choices remain.  The smectic may go directly to the
normal solid in a first order transition which breaks the
translational symmetry and restores the $U(1)$ invariance
simultaneously, as shown in Fig.\ref{sbfig}(b).  The last possibility,
illustrated in Fig.\ref{sbfig}(c), is that of an intermediate supersolid
phase between the smectic and the interstitial--free solid.  In this
case both the low temperature phase transitions may be second order.
The supersolid--solid critical behavior is described in
Ref.\onlinecite{FNF}.

\begin{figure}
\caption{Three routes of symmetry breaking at low temperatures
connecting the smectic phase to the vortex solid.  Choice (a) is ruled
out on physical grounds.  Note that $\langle \rho_\perp \rangle =
\langle \Phi \rangle \neq 0$ in all cases.}
\label{sbfig}
\end{figure}

The smectic--supersolid transition is once again a freezing transition
at a single wavevector, and is potentially describable by a Landau
theory like Eq.\ref{criticalfreeenergy}.  Because the modulating
effect of the underlying crystal lattice is much weaker in the $x$
direction, we expect $g \approx 0$ is a good approximation in this
case, which leads to pure XY behavior.

\section{Incommensurate Phases}
\label{ICphases}

As is well known from the study of the sine--Gordon
model,\cite{Villainreview}\ a large incommensurability can be
compensated for by energetically favorable ``solitons'', or walls
across which $u \rightarrow u + s$ (see Fig.\ref{ickinkpicfig}).
Solitons begin to proliferate when their field energy per unit area
$\sigma_{\rm field} \sim -\kappa{\cal A}s$ exceeds their cost at zero
field, $\sigma_0
\sim \sqrt{\kappa\tilde{g}} s$ (estimated from
Eq.\ref{smecticfreeenergy}).

Physically, these solitons correspond to extra/missing flux line layers
and walls of aligned ``jogs'' for $\delta{\bf H}$ along
the $b$ and $c$ axes, respectively (see Fig.\ref{ickinkpicfig}).  In the
former case, this leads to an incommensurate smectic (IS) phase, whose
periodicity is no longer a simple multiple of $s$.  For $\delta{\bf H}
\parallel {\bf\hat{z}}$, the solitons induce an additional periodicity
along the $y$ axis.  This tilted smectic (TS) phase has long range
translational order in two directions.\cite{crystalnote}\  The
analogous tilted {\sl crystal} (TX) phase is qualitatively similar,
but has long range order in 3 directions.

For larger $H_c$, as the angle between the field and the CuO$_2$
layers becomes large, intrinsic pinning and anisotropy no longer favor
the smectic state.  As shown in Fig.\ref{phasediagramfig}, we
therefore expect the L--TS and TS--TX phase boundaries to merge in
this regime.  The direct L--TX transition is necessarily first order.

\begin{figure}
\caption{Kinked configurations (neglecting fluctuations and drawn
viewed along the $x$ axis) of the smectic layers for magnetic
field perturbations (a) along the $b$ axis, and (b) along the $c$ axis.}
\label{ickinkpicfig}
\end{figure}

In conventional CITs, entropic contributions generate additional
interactions between domain walls which actually dominate over the bare
energetic repulsions when the inter--soliton spacing $\ell \rightarrow
\infty$.  To estimate their magnitude here, we use the well known
logarithmic roughness of a 2d interface,\cite{roughening}\
\begin{equation}
\left\langle \left( h({\bf x}) - h({\bf 0}) \right)^2\right\rangle \sim \ln|x|,
\label{logroughness}
\end{equation}
where $h$ is the height of the interface and the coordinate ${\bf x}$
parameterizes its position in the base plane.  For solitons spaced by
$\ell$, collisions between neighbors generally occur only once $h \gtrsim
\ell$, so that the size of roughly independently fluctuating regions $x
\sim \exp(\ell^2)$.  The entropy loss due to this constraint scales with
the number of collisions $(L/x)^2$, so that the areal free energy cost
per wall is
\begin{equation}
f_{\rm coll.} = -T\Delta s_{\rm coll.} \sim Te^{-\ell^2}.
\label{fewall}
\end{equation}
Since the energetic interactions in the smectic scale exponentially
(like $\exp (-\ell/w)$) at long distances, the collision free energy
is actually negligible as $\ell \rightarrow \infty$, unlike the
situation for lines in $1+1$ dimensions.\cite{Villainreview}\ The free
energy density in the incommensurate phases is thus
\begin{equation}
f_{\rm soliton} \sim -{{|\sigma|} \over l} + {\Delta \over
l}e^{-l/w},
\label{solitonfreeenergy}
\end{equation}
where $\sigma \equiv \sigma_{\rm field}+\sigma_0 <0$ is the total
areal free energy of the soliton; $\Delta$ and $w$ set the energy and
length scales of the soliton interactions.  At low temperatures we expect
$w \sim \lambda$ and $\Delta \sim \epsilon_0/a_x$, while near $T_s$,
Eq.\ref{smecticfreeenergy}\ gives $w \sim \sqrt{\kappa/\tilde{g}}s$ (c.f.
Eq.\ref{dwenergy}) and $\Delta \sim \tilde{g}\ell$.  Minimizing
Eq.\ref{solitonfreeenergy}\ gives a soliton separation $l
\sim w\ln(\Delta/|\sigma|)$ near the CIT.

In the TS phase, net vortex motion along the $c$ axis occurs by
sliding soliton walls along the $b$ direction.  The resulting electric
field is proportional to $J$ and the soliton density, leading to an
additional contribution to the resistivity which vanishes at the CIT
like $\rho_{xx}^{\rm soliton} \sim \rho_0^{\rm soliton}/\ln(\Delta/|\sigma|)$.

A single soliton wall in the IS phase, because it is parallel to the
CuO$_2$ layers, experiences a periodic potential along $z$.  From
studies of the roughening transition,\cite{roughening}\ it is known
that such a periodically pinned wall may be in either a rough or
smooth phase.  If the walls are individually smooth, thermal
fluctuations are negligible, and the assembly of solitons is well
described by an effectively one--dimensional elastic chain in a
periodic potential.\cite{Villainreview}\ Because they are pinned
separately into minima of the potential, they {\sl do not} contribute
to $\rho_{xx}$.  If they are rough, they wander logarithmically and
eventually interact with their neighbors.  The appropriate
coarse--grained description beyond this interacting length scale is an
elastic stack of domain walls.  The configuration of such a stack is
described by a second displacement field $u_{\rm dw}$, with a free
energy of the same form as  Eq.\ref{smecticfreeenergy} (but with
different values of $\kappa$ and $\tilde{g}$).  The statistical
mechanics for the $u_{\rm dw}$ field is thus equivalent to that of the
original $u$ variable.  The preceding analysis must then be repeated
within the new effective free energy.

Because of the aforementioned complexity of the one--dimensional
problem, we have not
attempted to determine the true long distance behavior of the soliton
array in the IS phase.  Because the permeation mode in the
commensurate smectic already provides a finite tilt modulus and
non--zero resistivity, however, we expect that these more subtle
effects will have only weak experimental implications.

As the temperature is increased within the IS or TS phases, the system
melts into the liquid.  To study such transitions, we
perform the dilation $\Phi \rightarrow \Phi \exp(i{\bf A\cdot r})$.
Only the $g$ term is not invariant under such a gauge--like
transformation.  It becomes oscillatory and therefore does not
contribute to the critical behavior at long wavelengths.  The IS--L
and TS--L phase transitions are thus XY--like.

The shape of the CIT phase boundary is of particular experimental
interest.  In the mean field regime, this is obtained from the
condition $\sigma=0$ as $\delta H \sim |r|^\Upsilon$, with
$\Upsilon_{\rm MF} = (m-2)/4$.  By the usual Ginzburg
criterion, mean field theory breaks down for $|r|
\lesssim (k_{\rm B}Tv/K^{3/2})^2$.  To determine the shape of the
 phase boundary in this critical regime, we follow the RG flows out of
the critical region and repeat the preceding analysis with the
renormalized couplings determined by matching when $|r|$ is order one.
Then $\delta H_{\rm R} \sim \xi^{\lambda_{\rm H}}\delta H$ and $g_{\rm
R} \sim
\xi^{\lambda_m}g$, with $\xi \sim |r|^{-\nu}$.  Rotational invariance
at the rescaled fixed point ($g=0$) implies that the field exponent is {\sl
exactly} $\lambda_{\rm H} = 1$ (see appendix \ref{rotinvappendix}).  Using
these
renormalized quantities, we find
\begin{equation}
\Upsilon_{\rm crit.} = (|\lambda_m|
+ 2)\nu/2 \approx 4.9 - 7.2,
\label{equationX}
\end{equation}
for the fields used in Ref.\onlinecite{Kwok}.

The IS--L and TS--L phase boundaries are non--singular and are
determined locally by the smooth $\delta {\bf H}$
dependence of $r$.  In particular, for small $H_c$, the TS--L phase
critical temperature is
\begin{equation}
T(H_c) = T_s - {{r''} \over 2r'}H_c^2,
\label{TSLboundary}
\end{equation}
where $r' \equiv \partial r/\partial T|_{T=T_s,\delta H=0}$.

\section{Influence of Disorder}
\label{disordersection}

Lastly, we consider the effects of weak point disorder, which couples to
the density of vortices according to
\begin{equation}
F_d =\int \! d^3\!{\bf r} V_d({\bf r}) n({\bf r}),
\label{disorder}
\end{equation}
where $V_d({\bf r})$ is a random potential, which, for point impurities,
is short range correlated in space and narrowly (e.g. Gaussian)
distributed at each point.  Using Eq.\ref{densitywave}, $F_d$ can be
rewritten, up to less relevant terms, as
\begin{equation}
F_d = \int \! d^3\!{\bf r} V_d({\bf r}){\rm Re} \{ \Phi({\bf r})e^{iqz} \},
\label{fdphi}
\end{equation}
in the smectic phase and in the liquid sufficiently near $T_s$.  To bring
Eq.\ref{fdphi}\ into a more standard form, we define a complex random
field $\tilde{V}_d \equiv V_d e^{iqz}$, in terms of which
\begin{equation}
F_d = \int \! d^3\!{\bf r} {1 \over 2}\left( \tilde{V}_d^* \Phi +
\tilde{V}_d \Phi^* \right).
\label{rfpert}
\end{equation}
Because of the oscillatory $e^{iqz}$ factor, $\tilde{V}_d$ and
$\tilde{V}^*_d$ are essentially uncorrelated at long wavelengths.
Eq.\ref{rfpert}\ is the simplest ``random field'' XY perturbation of
Eq.\ref{criticalfreeenergy}, and the resulting model is known in
statistical mechanics as a random field XY model with an $m$--fold
symmetry breaking term.

Before discussing the critical behavior of such a theory, it is natural
to consider the effect upon the ordered state.  In the smectic phase,
using $\Phi = \sqrt{|r|/v}\exp(2\pi i u/a)$, Eq.\ref{rfpert}\ becomes
\begin{equation}
F_d = \int \! d^3\!{\bf r} {1 \over 2} \sqrt{{|r| \over v}}\left(
\tilde{V}_d^* e^{2\pi i u/a} + \tilde{V}_d e^{-2\pi i u/a} \right).
\label{rfsmectic}
\end{equation}
If the disorder is weak relative to the periodic potential (i.e.
$|\tilde{V}_d| \ll \tilde{g}$), it is naively justified to replace the
cosine in Eq.\ref{smecticfreeenergy}\ by the ``mass'' term
$(2\pi/s)^2\tilde{g}u^2/2$.  Such a mass term gives a large penalty for
excursions of the layers with $u \gtrsim s$, so the randomness in
$F_d$ appears
irrelevant.

By ignoring the periodicity of the cosine, the above approach does not
consider the possibility of disorder--induced solitons.  To study the
stability of the smectic to such topological defects, consider a
region of size $L$ in which the displacement field $u$ is shifted by
$s$, so as not to incur any bulk energy cost from the intrinsic
pinning.  Within this region, $F_d$ contributes an energy of random
sign of order $|V_d|L^{d/2}$ in $d$ dimensions.  On the boundary of
the region, however, the cosine does contribute, costing an energy
$\sim \tilde{g}L^{d-1}$.  For $d>2$, the boundary energy grows more
rapidly with $L$, and the net energy is always positive, provided
$\tilde{g} > |V_d|s^{1-d/2}$.  Thus we see that the smectic phase
remains {\sl stable} to weak disorder, even once solitons are taken into
account.

Note that this result is in strong contrast to the Larkin--Ovchinikov
argument that the Abrikosov lattice is unstable to arbitrarily weak
pinning.\cite{LO}\  Physically, the instability is prevented, at least for
weak disorder, by the periodic pinning potential which increases the
stiffness of the smectic displacement field.  The result can be understood,
however, on more general symmetry grounds.  For non--zero $g$, the system
does not have a true continuous translational symmetry in the $z$
direction, but only the discrete symmetry under translations by $s$.
Because the symmetry is discrete, there is no Goldstone mode in the
ordered (smectic phase) -- i.e. phonons are massive.  It is now
well known that for random field models with discrete
symmetries (e.g. the random field Ising model, to which our model
corresponds when $m=2$), the ordered phase survives above two
dimensions.\cite{Imbrie}\  Indeed, the argument given above for stability
against droplet solitons is a restatement of the Imry--Ma argument first
used for the random field Ising model.\cite{ImryMa}\

In the incommensurate (IS and TS) phases, where the periodic pinning
$\tilde{g}$ is effectively zero, the Imry--Ma argument no longer
applies.  In these phases, the original Larkin--Ovchinikov picture
holds, and the distortions in $|u({\bf r})|^2$ must grow on long
length scales, destroying the long range translational order of the
layers.  The nature of the resulting phase is unclear: it may be a
``smectic glass'', analogous to the proposed vortex glass phase for
more isotropic systems, or it may simply be a strongly correlated
liquid, with slow relaxation times.  The same considerations hold for
the more ordered phases at low temperatures, since the translational
order along the $x$ axis lacks the intrinsic pinning required to prevent
the Larkin--Ovchinikov instability.

Turning to the critical behavior of the L--S transition, the analysis
becomes more subtle.  The random field perturbation in
Eq.\ref{rfpert}\ is a relevant perturbation at the XY fixed point (and
indeed at any $O(n)$ fixed point), so the critical behavior is
certainly altered.  Naively, a $6-\epsilon$ expansion may be made for
the critical behavior of the $O(n)$ random field model, which has a
zero temperature critical point.\cite{RFepsilon}\  Within such a
perturbative expansion near $6$ dimensions, the symmetry--breaking
term appears irrelevant.  There are two potential problems with this
approach.  Firstly, if the symmetry--breaking term indeed remains
irrelevant at the new critical point, it would be an example of a
three dimensional random field XY critical point.  The random field XY
model, however, because it has a continuous symmetry, does not even
have a stable ordered phase in three dimensions.  Although there may
not be an obvious contradiction involved in this scenario, the
physical meaning is certainly unclear.  One possible resolution is
that the symmetry--breaking term becomes relevant at some higher
dimension (greater than 4) .  The second problem is the status of the
$6-\epsilon$ expansion itself, which has been proven to break down, at
least via non--perturbative corrections (but possibly more strongly)
for the case of the random field Ising model.\cite{DSFrandom}\  The
consistency of the $6-\epsilon$ expansion, even perturbatively, has not
yet been determined.  Regardless of the success or failure of this
theoretical approach, experimental work on random field Ising systems
has demonstrated the subtle types of behavior possible for such zero
temperature critical points.  Fortunately, as discussed in sections
\ref{staticbehaviorsection}--\ref{supersolidsection}, the
supersolid nature of the smectic (with the relatively fast permeation
mode) implies that slow dynamics for the smectic ordering will not
have a strong impact on transport and magnetization experiments.

Finally, consider the L--IS and L--TS transitions in the presence of
disorder.  Disorder strongly effects the behavior at such CITs, because it
modifies the wandering of a single domain wall.\cite{KardarNelson,NL}\
Fortunately, the effects of random field disorder on a single interface
are known exactly.\cite{GM,BF}\  In contrast to Eq.\ref{logroughness}, the
height fluctuations of the interface grow like
\begin{equation}
\overline{\left\langle \left( h({\bf x}) - h({\bf 0})
\right)^2\right\rangle} \sim |x|^{2\zeta},
\label{randomroughness}
\end{equation}
where $\zeta = (5-d)/3 = 2/3$ in three dimensions.  Also unlike the pure
interface, the free energy fluctuations within a region grow with length
scale, so that the cost per collision scales like $|x|^\theta$, where
$\theta = d-3+2\zeta = 4/3$.  The areal collision free energy per wall is
thus
\begin{equation}
f_{\rm coll.} \sim 1/\ell^{(2-\theta)/\zeta} \sim 1/\ell.
\label{randomfcoll}
\end{equation}
This dominates over exponential energetic interactions for large $\ell$.
The full soliton free energy is thus
\begin{equation}
f_{\rm soliton} \sim -{{|\sigma|} \over l} + {{\Delta_d} \over \ell^2},
\label{rsolitonfreeenergy}
\end{equation}
where $\Delta_d$ measures the strength of the disorder--induced collision
interactions.  Minimization of Eq.\ref{rsolitonfreeenergy}\ gives $\ell
\sim 1/|\sigma|$.

\section{Conclusions and Applications to Helium Films}
\label{conclusionsection}

We have studied the behavior of vortex arrays subjected to a
one--dimensional periodic potential transverse to the magnetic field.
Such a potential, which is induced by the layered structure of the
high temperature copper oxide superconductors for fields oriented in
the a--b plane, favors an intermediate smectic phase between the
vortex lattice and flux liquid.  The commensurate smectic state is
supersolid, and has in consequence a nonzero finite resistivity and
tilt modulus, {\sl despite} being pinned by the periodic potential.
Including incommensurability effects leads to the rich phase diagrams
of Fig.\ref{phasediagramfig}.  The experimental signature of the
smectic is the appearance of Bragg peaks in the structure function
along a single ordering axis interleaving the trivial peaks induced by
the layering.  We also expect a greatly reduced resistivity for
currents transverse to both the layering and magnetic field axes, and
a cusped phase boundary describing the response to small fields
perpendicular to the layering direction.  The qualitative behavior of
the tilt modulus and resistivity near the liquid--to--smectic
transition in a field aligned perfectly with the ab--plane is shown in
Fig.\ref{qualfig}.

\begin{figure}
\caption{Tilt modulus and resistivity at $T_s$ for a perfectly aligned
commensurate field.  Not to scale: the resistivity could appear to
drop to zero and the tilt modulus could appear to diverge to infinity
in all but the most precise experiments.  The open circles denote
$|T-T_s|^{1-\alpha}$ singularities, where $\alpha$ is the specific
heat exponent.}
\label{qualfig}
\end{figure}

Using the boson mapping,\cite{NelsonSeung,duality}\ these results can
be extended to real two dimensional quantum mechanical bosons at zero
temperature.  The analogous quantum smectic phase might be studied in
helium on a periodically ruled substrate.  Such a substrate might be
approximated by crystalline facets exposing a periodic array of
rectangular unit cells with large aspect ratio.  There are, however, a
number of difficulties inherent in this extension.  In particular, the
interaction between helium atoms is not purely repulsive; it is
reasonably well described by a Lennard--Jones potential with a minimum
at an interatomic spacing of a few Angstroms.  To obtain a substrate
with a small enough period to affect the physics on these length
scales is an experimental challenge.  Were these interactions purely
repulsive, one could probably overcome this difficulty by working with
a dilute system.  With an attractive tail to the potential, however, a
low density helium film would likely phase separate into helium rich
and helium poor regions, making intermediate densities inaccessible.
Appropriate experimental conditions may nevertheless be achievable for
small values of $m$ (the number of periods of the potential per period
of the smectic density wave).  The depth of the minimum in the
effective pair potential, moreover, could be reduced somewhat by a
careful choice of substrate.

The case $m=2$ has been explored numerically in a Bose Hubbard model
in Ref.\onlinecite{Zimanyi}.  These authors indeed find a smectic
phase, which they denote a ``striped solid'', with order in reciprocal
space at ${\bf q} = (\pi,0)$.  Their results for the structure
function and superfluid density appear to be in good agreement with
the predictions of section \ref{smecticsection}\ (see in particular
their Fig.11.  The superfluid density of the boson system maps onto
the inverse tilt modulus $c_{44}^{-1}$ of the flux
lines.\cite{BGtheory}).  Unfortunately, a detailed comparison of the
critical behavior with the theory is beyond the resolution of the
available data.

At a more general level, the liquid--smectic transition treated here
appears to be the only known case of continuous quantum freezing
in $2+1$ dimensions.  The smectic phase, moreover, is
perhaps the simplest example of a quantum phase intermediate between
solid and liquid.  The techniques developed here may be useful in
understanding other quantum phases of mixed liquid/solid character.
One particularly intriguing example is the ``Hall solid'' proposed in
Ref.\onlinecite{Hallsolid}.  Through the Chern--Simons
mapping,\cite{ChernSimons}\ it can be related to a supersolid phase of
composite bosons (electrons plus flux tubes).\cite{Bunpub}\ This
phase, an analogous ``Hall smectic'' and ``Hall hexatic,'' and the
modifications of the current theory to account for the long--range
Coulomb and Chern--Simons interactions are discussed in
Ref.\onlinecite{Bunpub}.

\acknowledgements It is a pleasure to acknowledge discussions with
George Crabtree, Daniel Fisher, Matthew Fisher, Randall Kamien, Wai
Kwok, Leo Radzihovsky, and John Reppy.  This research was supported by
the National Science Foundation, through grant No. DMR94--17047 and in
part through the MRSEC program via grant DMR4--9400396.  L.B.'s work
was supported at the Institute for Theoretical Physics by grant
no. PHY89--04035.

\appendix

\section{Derivation of Hydrodynamic Equations}
\label{dynamicsappendix}

The continuity equation (Eq.\ref{continuity}) follows from
differentiation of Eqs.\ref{ndef}.  For example,
\begin{eqnarray}
\dot{n} & = & -\sum_i \left[\delta'(x-x_i)\delta(z-z_i)\dot{x}_i +
\delta(x-x_i)\delta'(z-z_i)\dot{z}_i \right] \nonumber \\
& = & -\bbox{\nabla}_\perp \cdot \sum_i \delta({\bf r}_\perp - {\bf
r}_{\perp i}) {\bf \dot{r}}_{\perp i} = - \bbox{\nabla}_\perp \cdot
{\bf j}_{v},
\label{contderiv}
\end{eqnarray}
where
\begin{equation}
{\bf j}_v \equiv \sum_i \delta({\bf r}_\perp - {\bf r}_{\perp i}){\bf
\dot{r}}_{\perp i}.
\label{jvdef}
\end{equation}
Eq.\ref{taucontinuity}\ is derived analogously, giving the tangent
current tensor
\begin{equation}
j_{\beta\alpha} \equiv \sum_i \delta({\bf r}_\perp - {\bf r}_{\perp
i}) \left( {{\partial x_i^\beta} \over {\partial t}}{{\partial
x_i^\alpha} \over {\partial y}} - {{\partial x_i^\alpha} \over
{\partial t}}{{\partial x_i^\beta} \over {\partial y}}\right).
\label{jtdef}
\end{equation}

Eqs.\ref{jvdef}\ and \ref{jtdef}\ are completely general, and do not
depend upon the detailed dynamics of the vortex system.  This
additional physics
is included in the constitutive equations (Eqs.\ref{constitutive}\ and
\ref{tauconstitutive}).  To derive them, we need the equation of
motion, Eq.\ref{FLeom}.  Inserting this into Eq.\ref{jvdef}\ gives
\begin{equation}
\Gamma {\bf j}_v = - \sum_i \delta({\bf r}_\perp - {\bf r}_{\perp
i}){{\delta F} \over {\delta{\bf r}_{\perp i}(y)}} + n{\bf f}.
\label{jvintermediate}
\end{equation}
The functional derivative with respect to ${\bf r}_{\perp i}$ can be
transformed via the chain rule
\begin{equation}
\delta F = \int d^3{\bf r} \left[ {{\delta F} \over {\delta n({\bf r})}}
\delta n({\bf r}) + {{\delta F} \over {\delta \bbox{\tau}({\bf r})}}
\cdot \delta\bbox{\tau}({\bf r})\right],
\label{chainrule}
\end{equation}
where the variations $\delta n$ and $\delta\bbox{\tau}$ are
\begin{eqnarray}
\delta n({\bf r}) & = & -\bbox{\nabla}_\perp\cdot \sum_i \delta({\bf
r}_\perp - {\bf r}_{\perp i}) \delta{\bf r}_{\perp i}(y), \nonumber\\
\delta\bbox{\tau}({\bf r}) & = & -\partial_\alpha \sum_i \delta({\bf
r}_\perp - {\bf r}_{\perp i}) {{d{\bf r}_{\perp i}} \over {dy}} \delta
x_i^\alpha(y) \nonumber \\
& & + \sum_i \delta({\bf r}_\perp - {\bf r}_{\perp i}) {{d\delta{\bf
r}_{\perp i}} \over {dy}}.
\label{variations}
\end{eqnarray}
Substituting Eqs.\ref{chainrule}\ and \ref{variations} into
Eq.\ref{jvintermediate}\ then gives Eq.\ref{constitutive}.  The
constitutive equation for $j_{\beta\alpha}$ is derived analogously, by
premultiplying Eq.\ref{FLeom}\ with $\partial x_i^\beta/\partial y$
and carrying out the same steps as before.

\section{Landau Theory of the Smectic to Crystal Transition}
\label{sxappendix}

We assume that ${\bf H}$ is in the ab--plane and commensurate smectic
order is already well established, and ask how a {\sl
two}--dimensional vortex modulation then arises at a lower
temperature.  The modulated vortex density now takes the form
\begin{eqnarray}
n({\bf r}) & = & n_0 {\rm Re}\bigg\{ 1 + \Phi({\bf r})e^{-iqz} +
\psi_1({\bf r})e^{-i {\bf G_1 \cdot r}} \nonumber \\
& & + \psi_2({\bf r})e^{-i {\bf
G_2 \cdot r}} \bigg\},
\label{X1}
\end{eqnarray}
where $\Phi({\bf r})$ is the (large) smectic order parameter, and
${\bf G}_1$ and ${\bf G}_2$ are reciprocal lattice vectors lying in
the $(x-z)$ plane with $G_{1x} = - G_{2x} \neq 0$ satisfying
\begin{equation}
q{\bf \hat{z}} + {\bf G_1} + {\bf G_2} = \bbox{0}.
\label{X2}
\end{equation}
The six vectors $\pm q{\bf \hat{z}}$, $\pm {\bf G_1}$, and $\pm {\bf
G_2}$ form a distorted hexagon of minimal reciprocal lattice vectors.
All other reciprocal lattice vectors in the crystalline phase are
linear combinations of this set, which reflects an anisotropic vortex
lattice in real space.  The corresponding set of reciprocal lattice
vectors for a {\sl square} lattice is illustrated in Fig.\ref{sffig}.

The complex amplitudes $\psi_1({\bf r})$ and $\psi_2({\bf r})$ are
small near the transition, and the Landau free energy difference
$\delta F$ between the smectic and crystalline phases takes the form
\begin{eqnarray}
\delta F & = & \int \! d^3{\bf r} \bigg[ {\tilde{K} \over
2}\left|\nabla\psi_1\right|^2 + {\tilde{K} \over
2}\left|\nabla\psi_2\right|^2 + \tilde{g} \bbox{\nabla}\psi_1 \cdot
\bbox{\nabla}\psi_2 \nonumber \\
& + & {r \over 2}(|\psi_1|^2 + |\psi_2|^2) +
\tilde{w}(\Phi\psi_1\psi_2 + \Phi^*\psi_1^*\psi_2^*) + \cdots \bigg].
\label{X3}
\end{eqnarray}
We have equated the coefficients of gradients in all three directions
for simplicity.  Within mean field theory, crystalline order can arise
via a continuous phase transition whenever $r<0$.  The neglected
higher order terms fix the magnitudes of $\psi_1$ and $\psi_2$,
$|\psi_1| = |\psi_2| \equiv \psi_0$ below the mean--field transition
temperature.  To study the true transition (which occurs for $r=r_c
<0$ due to thermal fluctuations), we set
\begin{equation}
\psi_1 = \psi_0 e^{i\theta_1({\bf r})}, \hspace{0.5 truein} \psi_2 =
\psi_0 e^{i\theta_2({\bf r})}.
\label{X4}
\end{equation}
Upon neglecting a constant, the free energy becomes
\begin{eqnarray}
\delta F & = & \int \! d^3{\bf r} \bigg[ {K \over 2}|\nabla\theta_1|^2
+ {K \over 2}|\nabla\theta_2|^2 \nonumber \\
& & + g \bbox{\nabla}\theta_1 \cdot \bbox{\nabla}\theta_2 + w
\cos(\theta_1 - \theta_2) \bigg],
\label{X5}
\end{eqnarray}
where $K= \tilde{K}\psi_0^2$, $g=\tilde{g}\psi_0^2$,
$w=\tilde{w}|\Phi|\psi_0^2$, and we have assumed the phase of the
smectic order parameter $\Phi$ is locked to zero by the periodic
pinning potential.  Eq.\ref{X5}\ represents two coupled XY models with
phases locked by the cosine.  This term forces $\theta_1 \approx
\theta_2 \equiv \theta$, and the phase transition falls in the
universality class of a three--dimensional XY model with effective
free energy
\begin{equation}
\delta F_{\rm XY} \approx (K+g)\int \! d^3{\bf r} |\nabla\theta|^2.
\label{X6}
\end{equation}

\section{Effect of Long Wavelength Fluctuations at the S--L Critical Point}
\label{murelevanceappendix}

To determine $\lambda_\gamma$, we assume the usual scaling form of the
free energy at the critical point,
\begin{equation}
f(r,\gamma) = \xi^{-d} g(\gamma\xi^{\lambda_\gamma}),
\label{freeenergyscalingform}
\end{equation}
where $g$ is an unknown scaling function.  Using $\xi \sim r^{-\nu}$ and
$2-\alpha=d\nu$,
Eq.\ref{freeenergyscalingform}\ gives
\begin{equation}
\left.{{\partial^2 f} \over {\partial\gamma^2}}\right|_{\gamma = 0} =
\xi^{2\lambda_\gamma -d}g(0) \sim r^{2-\alpha-2\lambda_\gamma\nu}.
\label{scalingprediction}
\end{equation}
The same quantity can be calculated directly, however, by differentiating the
partition function to obtain
\begin{equation}
\left.{{\partial^2 f} \over {\partial\gamma^2}}\right|_{\gamma = 0} = \int
d^3{\bf x} \langle \bbox{\nabla}_\perp \!\cdot\! {\bf w}({\bf x})
\bbox{\nabla}_\perp \!\cdot\! {\bf w}({\bf 0}) \rangle \langle E({\bf x})
E({\bf 0}) \rangle,
\label{zdiff}
\end{equation}
where the ``energy'' operator $E({\bf x}) \equiv |\Phi({\bf x})|^2$.  The
angular brackets indicate expectation values in the decoupled theories.
The energy--energy correlations take the scaling form
\begin{equation}
\langle E({\bf x})E({\bf 0})\rangle \sim r^{2(1-\alpha)}h(|{\bf x}|/\xi),
\label{Ecorrelations}
\end{equation}
where $h(\chi)$ is another unknown scaling function.  The
long--wavelength fluctuations, determined from Eq.\ref{wfreeenergy},
are
\begin{eqnarray}
\langle & &\bbox{\nabla}_\perp \!\cdot\! {\bf w}({\bf x})
\bbox{\nabla}_\perp \!\cdot\! {\bf w}({\bf 0}) \rangle  = n_0^2 k_{\rm
B}T \times
\nonumber \\
& & \int \!
{{d^3{\bf q}} \over {(2\pi)^3}} {{c_{44,\perp}q_x^2 +
c_{44,\parallel}q_z^2} \over {c_{11}(c_{44,\perp}q_x^2 +
c_{44,\parallel}q_z^2) + c_{44,\perp}c_{44,\parallel}q_y^2}} e^{i{\bf
q}\cdot {\bf x}}.
\label{wwcorrelations}
\end{eqnarray}
Inserting Eqs.\ref{Ecorrelations}\ and \ref{wwcorrelations}\ into
Eq.\ref{zdiff}\ and changing variables ${\bf x} \rightarrow \xi{\bf
x}$ gives the scaling
\begin{equation}
\left.{{\partial^2 f} \over {\partial\gamma^2}}\right|_{\gamma = 0} \sim
r^{2(1-\alpha)}.
\label{zdiffscaling}
\end{equation}
Comparison with Eq.\ref{scalingprediction}\ then gives the desired
result $\lambda_\gamma = \alpha/2\nu$.

\section{Equations of Critical Dynamics}
\label{eomappendix}

To derive the critical equations of motion, we assume that important
fluctuations occur only near ${\bf q}=\bbox{0}$ and ${\bf q} = q{\bf\hat{z}}$.
Including the ${\bf q}=\bbox{0}$ modes in Eq.\ref{densitywave}\ gives
\begin{equation}
n({\bf r}) = n_0 \left[1 + {\rm Re} \Phi e^{-iqz})  \right] + \delta n.
\label{densityexp}
\end{equation}
It is sufficient to keep only the small ${\bf q}$ parts of the tangent
field.\cite{BNunpublished}\ The finite $q_z$ modulation of the density
induces a modulation of the vortex current,
\begin{equation}
{\bf j}_{v} = {\bf j}_{v, u} + {\rm Re} {\bf j}_{v, s}e^{-iqz},
\label{currentdecomp}
\end{equation}
where ${\bf j}_{v, u}$ and ${\bf j}_{v, s}$ are the (slowly varying)
uniform and smectic (i.e. periodic) components of the vortex current.
Inserting
Eq.\ref{densityexp}\ into Eq.\ref{constitutive}\ and isolating the
parts proportional to $e^{-iqz}$ gives
\begin{eqnarray}
\Gamma{\bf j}_{v, s} & = & -4\left(1 + {{\delta n} \over n_0} \right)
\left( \bbox{\nabla}_\perp - iq{\bf\hat{z}}\right){{\delta F} \over
{\delta\Phi^*}} - n_0\Phi\bbox{\nabla}_\perp{{\delta F} \over
{\delta(\delta n)}} \nonumber \\
& & + n_0\Phi\partial_y{{\delta F} \over
{\delta\bbox{\tau}}} + n_0\Phi {\bf f} + n_0 {\bf f}_{\rm thermal}e^{iqz}
\label{jvs}
\end{eqnarray}
where the last term is included because the white noise ${\bf f}_{\rm
thermal}$ has Fourier components at all wavevectors.  Similarly, from
the uniform component, one finds
\begin{eqnarray}
\Gamma{\bf j}_{v, u} & = & -(n_0 + \delta n)\left[\bbox{\nabla}_\perp
{{\delta F} \over {\delta(\delta n)}} - \partial_y {{\delta F} \over
{\delta\bbox{\tau}}}\right] - \tau_\alpha\bbox{\nabla}_\perp {{\delta
F} \over {\delta\tau_\alpha}} \nonumber \\
& & - 2{\rm Re}\left[\Phi\left(\bbox{\nabla}_\perp +
iq{\bf\hat{z}}\right){{\delta F} \over {\delta\Phi}}\right] + (n_0 +
\delta n){\bf f}.
\label{jvu}
\end{eqnarray}
Using Eqs.\ref{currentdecomp}, the divergence of the current is
\begin{equation}
\bbox{\nabla}_\perp \cdot {\bf j}_v = \bbox{\nabla}_\perp \cdot {\bf
j}_{v ,u} + {\rm Re} (\bbox{\nabla}_\perp -iq{\bf\hat{z}}){\bf j}_{v,
s}e^{-iqz},
\label{cdiv}
\end{equation}
which leads to the two continuity equations
\begin{eqnarray}
\partial_t \delta n + \bbox{\nabla}_\perp\cdot {\bf j}_{v,u} & = & 0,
\label{uconst} \\
n_0\partial_t \Phi + (\bbox{\nabla}_\perp -iq{\bf\hat{z}})\cdot {\bf
j}_{v,s} & = & 0. \label{sconst}
\end{eqnarray}

Eqs.\ref{jvs}--\ref{sconst}\ completely specify the dynamics of the
density fluctuations $\delta n$ and $\Phi$.  Inserting Eq.\ref{jvs}\
into Eq.\ref{sconst}, and neglecting irrelevant couplings gives the
``model E''--like model of Eq.\ref{criticaldynamics}.

Eq.\ref{jvu}\ can, in principle, be used to study the effects of the
smectic ordering on the long--wavelength modes.  The most physical
application, however, is to determine the contribution of the smectic
degrees of freedom to the electric field.  The average bulk
(i.e. ${\bf q=0}$) field is $\langle {\cal E}_x \rangle =
\langle j_{v, u,z} \rangle \phi_0/c$.  In the spirit of the Landau
expansion, ${\bf q=0}$ component of Eq.\ref{jvu}\ can be reasonably
approximated by dropping terms with gradients of $\Phi$ (the leading
contributions from the first term involving $\delta n$ and
$\bbox{\tau}$ vanished automatically at ${\bf q=0}$ even without this
approximation), yielding
\begin{equation}
\Gamma \langle {\cal E}_c \rangle \approx n_0\phi_0 f_z/c +
{{iq\phi_0} \over c} \left[ \Phi^*{{\delta F} \over {\delta\Phi^*}} -
\Phi{{\delta F} \over {\delta\Phi}} \right].
\label{exintermediate}
\end{equation}
Recognizing the variation of the free energy on the right hand side of
Eq.\ref{exintermediate}, the substitution (c.f. Eq.\ref{criticaldynamics})
\begin{equation}
{{\delta F} \over {\delta\Phi^*}} = -{{\gamma_{\rm BS}} \over
{4q^2}}\partial_t\Phi + {{i\mu J_x} \over {4q^2}}\Phi,
\label{substitution}
\end{equation}
leads immediately to Eq.\ref{vortexcurrent}.

\section{Eigenvalue of ${\bf A}$}
\label{rotinvappendix}

The eigenvalue $\lambda_{\rm H}$ is determined completely by rotational
invariance.  To see this, consider Eq.\ref{criticalfreeenergy}\ at the
fixed point, i.e.  with $g=0$.  By making the transformation $\Phi
\rightarrow \Phi \exp(i{\bf A}\cdot{\bf r})$, the term involving ${\bf A}$
may be completely removed from $F$.  The free energy is thus independent of
${\bf A}$.  After integrating out short--wavelength modes, no dependence on
${\bf A}$ can appear, so precisely the same transformation must eliminate
the field dependence in the
renormalized free energy.  This operation is $\Phi \rightarrow \Phi
\exp(i{\bf A} \cdot {\bf r}\xi)$, using the rescaled ${\bf r}$, so the
renormalized vector potential must be
\begin{equation}
{\bf A}_{\rm R} = \xi {\bf A},
\label{rA}
\end{equation}
which implies $\lambda_{\rm H} =1$.


\begin{references}
\bibitem{cuprateref} In real cuprate materials, there is usually more
than one such plane per unit cell spacing in the $z$--direction.  In
this paper, we will usually ignore such complications, and model the
potential by a simple sinusoidal variation along $z$.
\bibitem{xdelocref} The line is of course still delocalized along the
$x$--direction, not shown in the figure.
\bibitem{localnote} Both the equal--$y$ interaction and the
wavevector--independent elastic moduli are appropriate in
situations when the curvature of the vortex lines is small on
the scale of $\lambda$.  See Appendix B of Ref.\onlinecite{BGtheory}.
\bibitem{derivationnote} The static equilibrium hydrodynamic equations
described here have be derived from the boson mapping in
Ref.\onlinecite{KN}.
\bibitem{smallmfoot} For $m<m_c$, symmetry breaking terms are {\sl
relevant}, and change the universality class of the transition.
The cases $m=2$ and $m=3$ correspond to Ising and three--state Potts
models, repspectively.
\bibitem{hdepnote} In fact, $K$ and $v$ and other higher
order coefficients also depend on $H_c$.  Their contributions to
$c_{44,\perp}$, are, however, small in the limit considered.
\bibitem{conserveddensitynote} Our problem differs from model E in that
the ``conserved density'' $\delta n$ is intimately connected via
Eq.\ref{constraint}\ to the tangent field $\tau$.
\bibitem{u1note} It may seem surprising that the $U(1)$ symmetry is
unbroken at zero temperature.  In the language of statistical
mechanics, this is because $\psi$ is actually a {\sl disorder}
operator, as can be seen by the duality relation with the original
Ginzburg--Landau order parameter $\Psi_{\rm GL}$.\cite{duality}\
\bibitem{crystalnote} Since the solitons are two--dimensional objects
embedded in three dimensions, dislocations cannot occur without
terminating the wall itself, which is presumably energetically
very costly.  A dislocation free stack of layers in three dimensions
automatically has long--range order, justifying the description as a
crystalline phase.
\end{references}
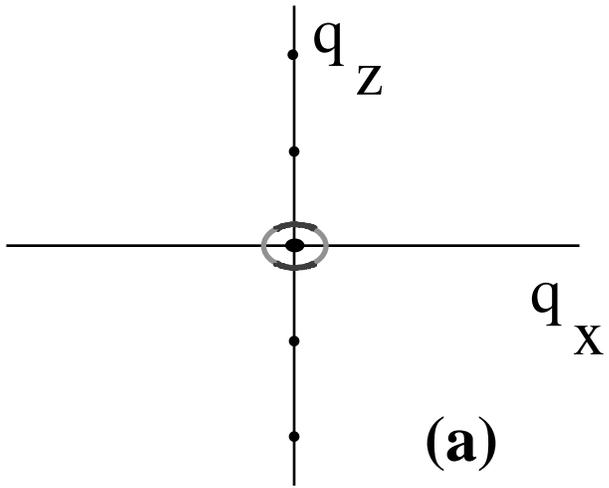
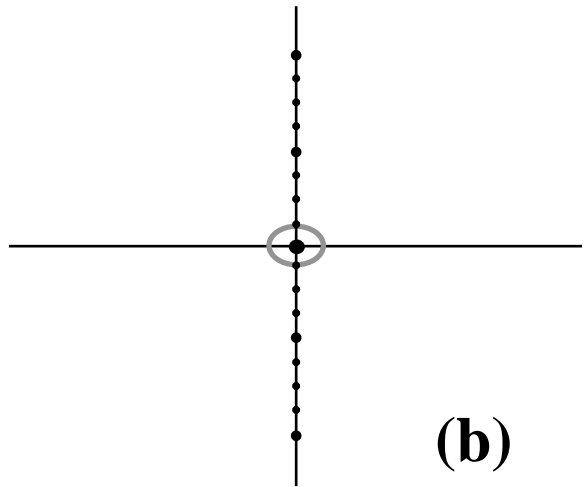
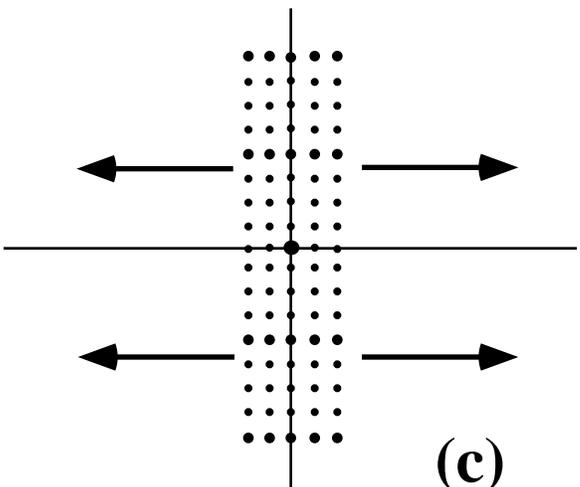
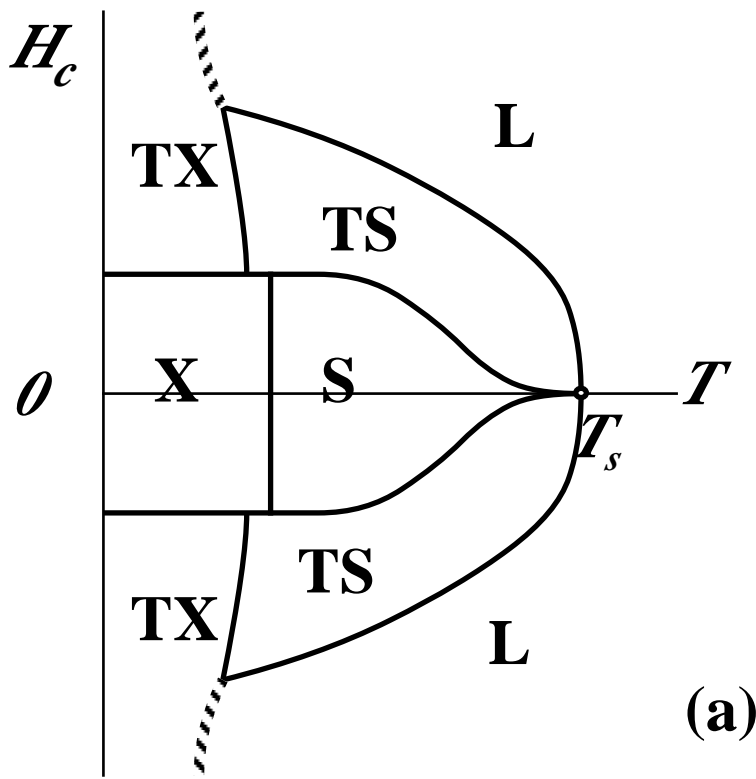
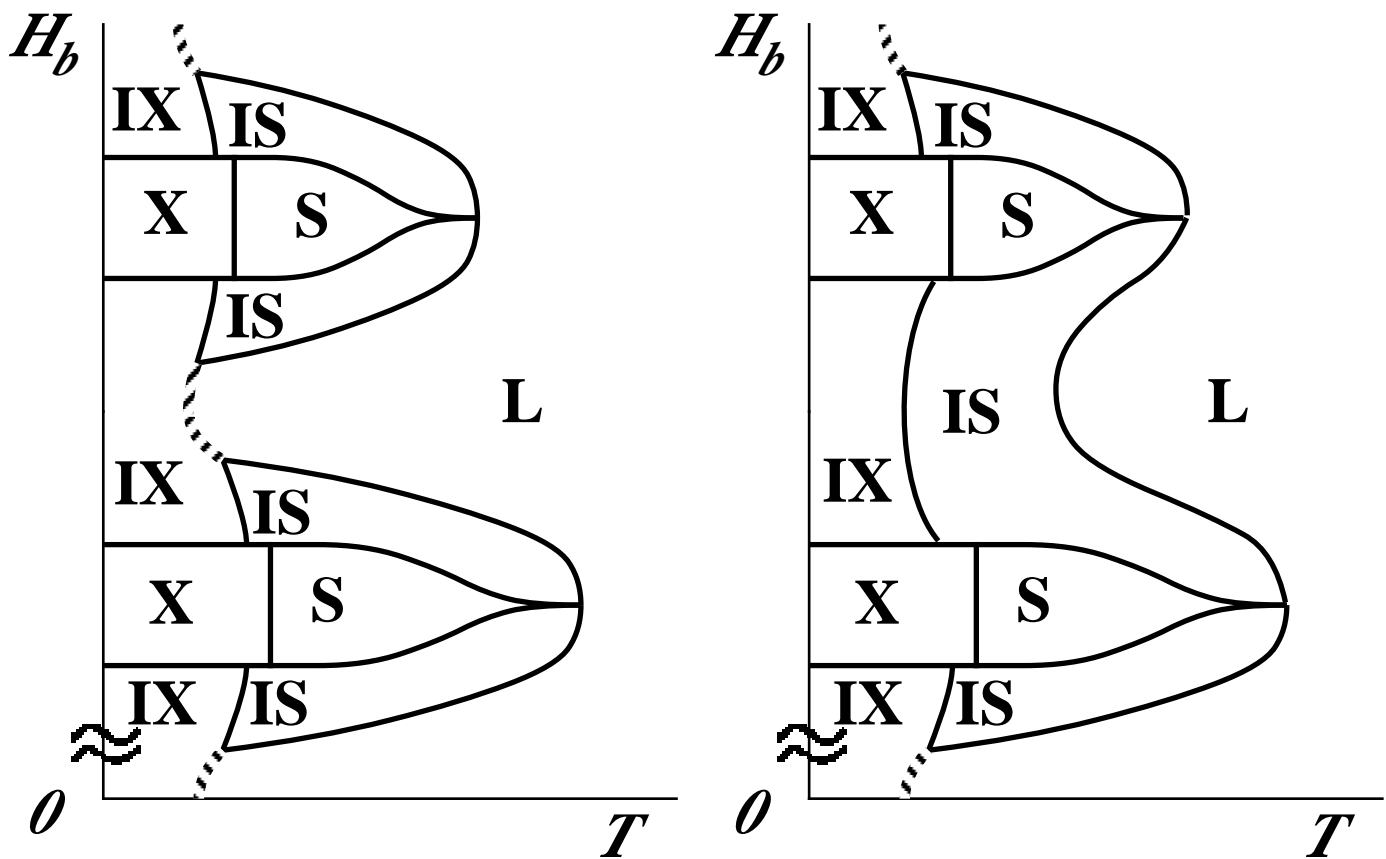
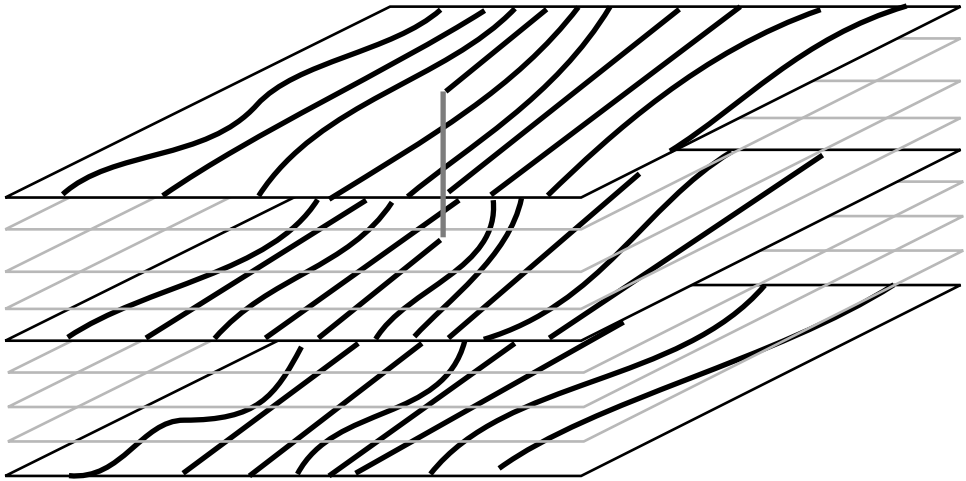
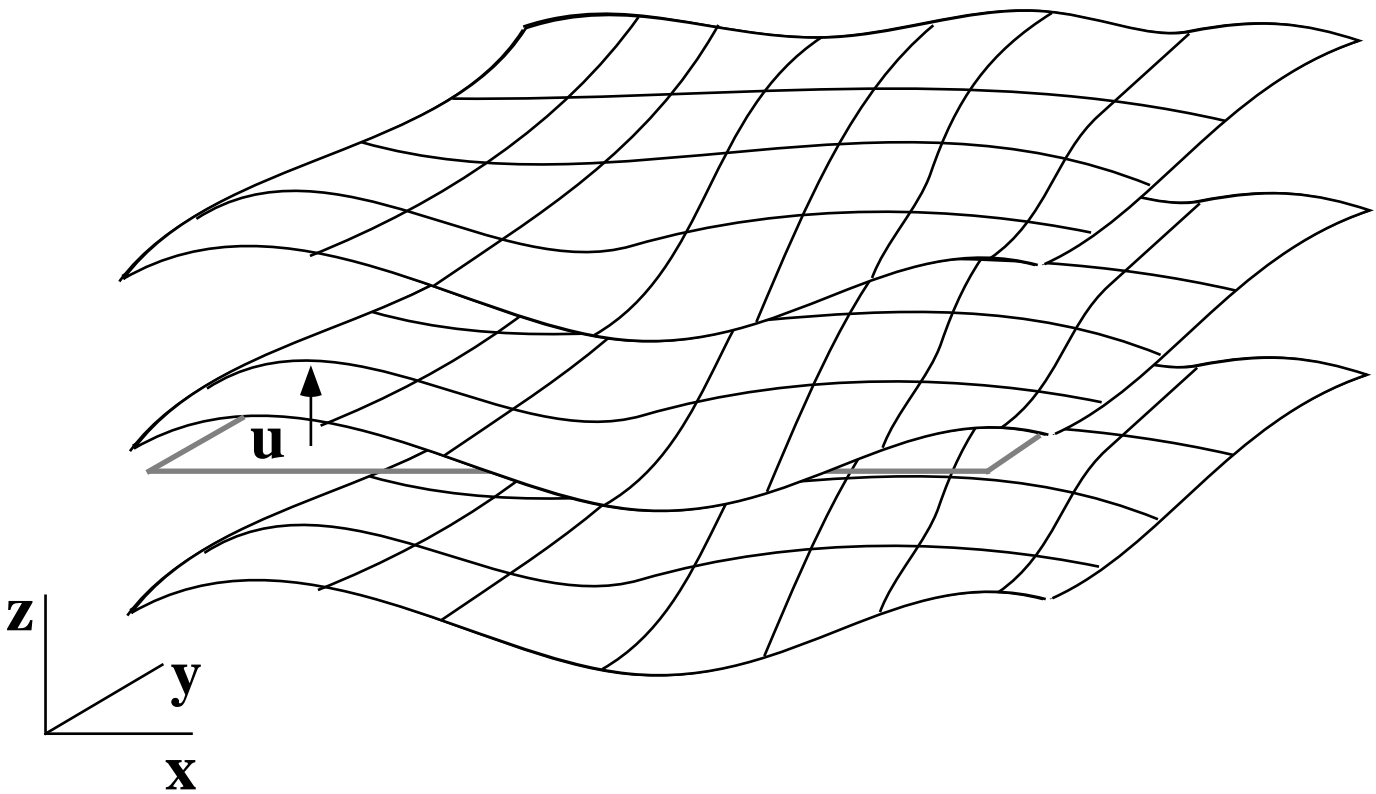
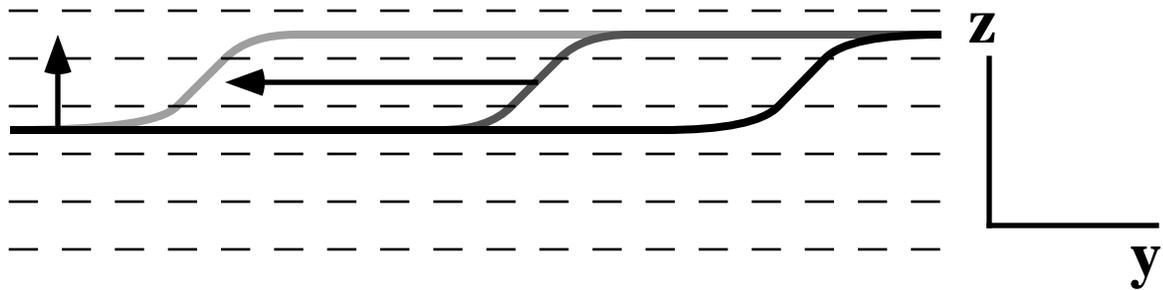
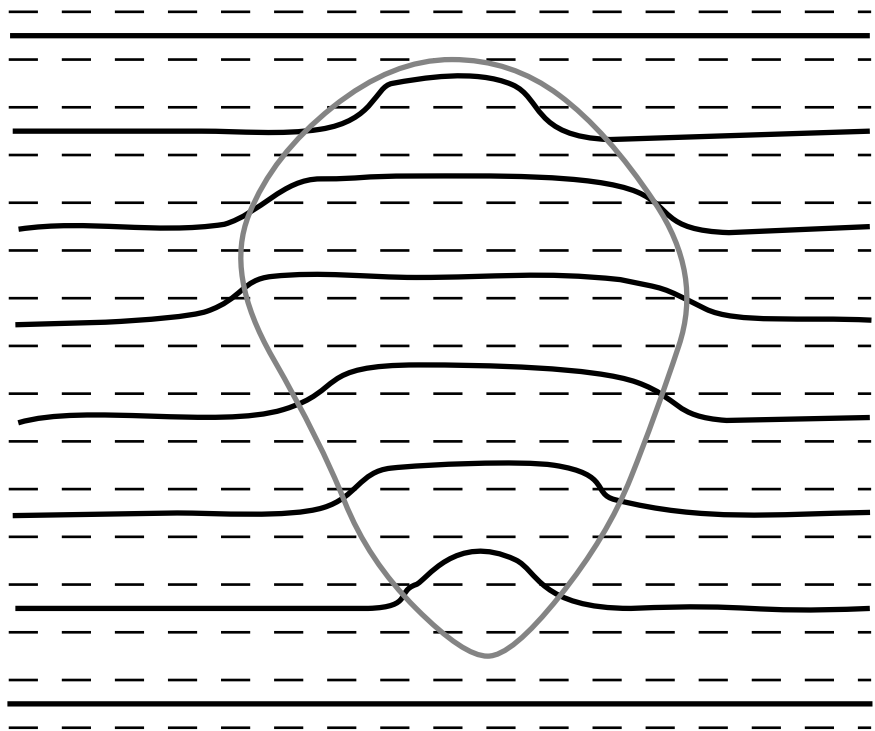
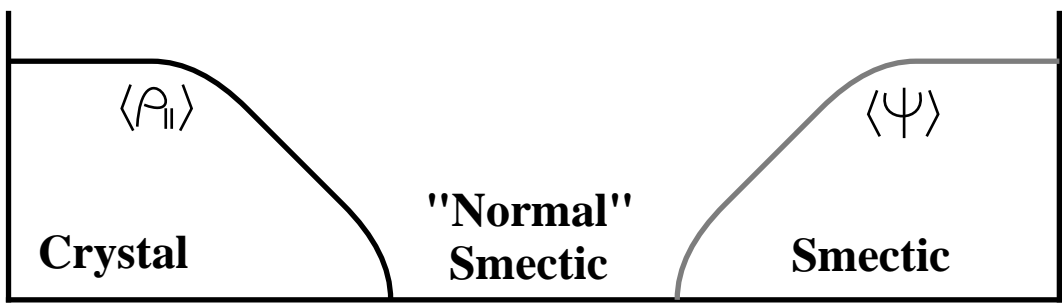
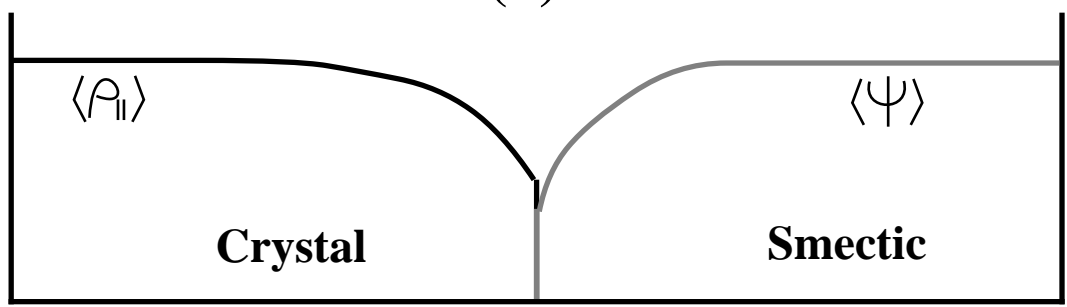
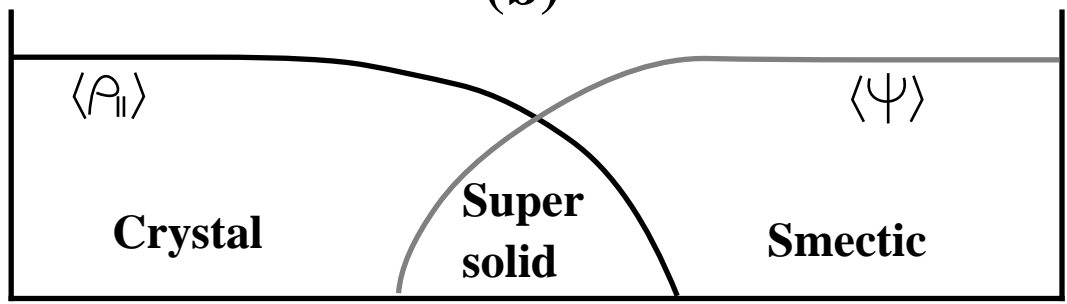
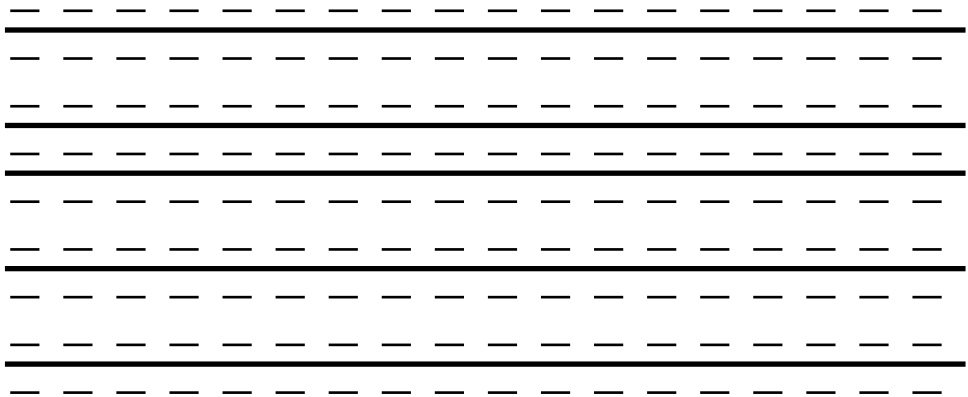
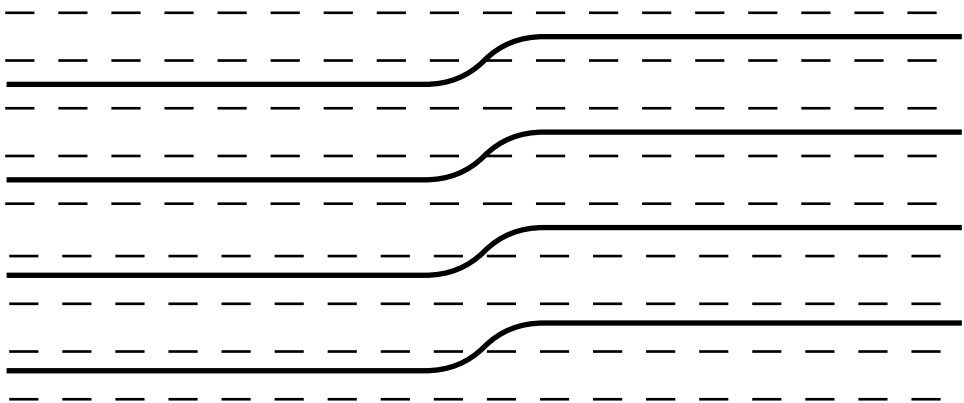
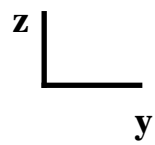
\end{document}